\documentclass[journal]{IEEEtran}
\usepackage[caption=false,font=footnotesize]{subfig}
\usepackage{bigints}
\usepackage{algorithm}
\usepackage{algpseudocode}
\usepackage{amsmath}
\usepackage{amsfonts}
\usepackage{amssymb}
\usepackage{amsthm}
\usepackage{mathtools}
\usepackage{color,soul}
\usepackage[detect-all,detect-weight=true]{siunitx}
\usepackage{booktabs}
\usepackage{pgfplots}
\usepackage{cite}
\usepackage{textcomp}
\usepackage[hidelinks]{hyperref}

\usepgfplotslibrary{fillbetween}
\usetikzlibrary{pgfplots.polar, pgfplots.fillbetween, matrix, shapes.geometric, arrows, patterns} 

\tikzstyle{input} = [rectangle, rounded corners, minimum width=2cm, minimum height=1cm, text centered, draw=black, fill=orange!30]
\tikzstyle{output} = [rectangle, rounded corners, minimum width=2cm, minimum height=1cm, text centered, draw=black, fill=blue!30]
\tikzstyle{table} = [diamond, minimum width=2cm, minimum height=1cm, text centered, draw=black, fill=green!30]

\pgfdeclarelayer{background}
\pgfsetlayers{background,main}

\DeclareMathOperator{\diag}{diag}

\DeclareMathOperator{\trace}{Tr}

\definecolor{lines-1}{RGB}{228,26,28}
\definecolor{lines-2}{RGB}{55,126,184}
\definecolor{lines-3}{RGB}{77,175,74}
\definecolor{lines-4}{RGB}{152,78,163}
\definecolor{lines-5}{RGB}{255,127,0}
\definecolor{lines-6}{RGB}{255,255,51}
\definecolor{lines-7}{RGB}{166,86,40}
\definecolor{lines-8}{RGB}{247,129,191}
\definecolor{lines-9}{RGB}{153,153,153}

\definecolor{verylightyellow}{RGB}{242,237,89}
\definecolor{verylightgray}{RGB}{200,200,200}
\definecolor{verylightred}{RGB}{255,191,191}
\definecolor{verylightgreen}{RGB}{117,234,123}

\pgfplotscreateplotcyclelist{myCycleList}{
	color=lines-1,thick,mark=o,mark size=3\\%
	color=lines-2,thick,mark=star,mark size=3\\%
	color=lines-3,thick,mark=square,mark size=3\\%
	color=lines-4,thick,mark=diamond,mark size=3\\%
	color=lines-5,thick,mark=triangle,mark size=3\\%
	color=lines-8,thick,mark=Mercedes star,mark size=3\\%
}

\pgfplotsset{
	compat=1.3,
	width =\columnwidth, 
	height=.7\columnwidth,
	ylabel absolute, ylabel style={yshift=-0.30cm},
	xlabel absolute, xlabel style={yshift=0.2cm},
	label style={font=\footnotesize},
	tick label style={font=\footnotesize},
	legend style={font=\footnotesize},
}

\DeclareMathOperator{\E}{\operatorname{\mathbb{E}}}
\DeclareMathOperator{\rank}{\operatorname{rank}}

\newtheorem{problem}{Problem}

\makeatletter
\newcommand{\pushright}[1]{\ifmeasuring@#1\else\omit\hfill$\displaystyle#1$\fi\ignorespaces}
\newcommand{\pushleft}[1]{\ifmeasuring@#1\else\omit$\displaystyle#1$\hfill\fi\ignorespaces}
\makeatother

\begin{document}
	
	\title{Optimal Precoders for Tracking \\the AoD and AoA of a mmWave Path}

	\author{
		Nil~Garcia~\IEEEmembership{Member,~IEEE}, Henk~Wymeersch~\IEEEmembership{Member,~IEEE}, Dirk~Slock~\IEEEmembership{Fellow,~IEEE}
		\thanks{N.~Garcia and H.~Wymeersch are with the Department of Signals and Systems, Chalmers University of Technology, Gothenburg, Sweden.
			D. Slock is with the Department of Communication Systems, EURECOM, Biot, France.
			This research was supported in part, by the EU HIGHTS project (High precision positioning for cooperative ITS applications) MG-3.5a-2014-636537, the VINNOVA COPPLAR project, funded under Strategic Vehicle Research and Innovation grant No.~2015-04849.}
	}
	
	\maketitle
	
	\begin{abstract}
		In millimeter-wave channels, most of the received energy is carried by a few paths. 
		Traditional precoders sweep the angle-of-departure (AoD) and angle-of-arrival (AoA) space with directional precoders to identify directions with largest power. Such precoders are heuristic and lead to sub-optimal AoD/AoA estimation. We derive optimal precoders, minimizing the Cram\'{e}r-Rao bound (CRB) of the AoD/AoA, assuming a fully digital architecture at the transmitter and spatial filtering of a single path. The precoders are found by solving a suitable convex optimization problem. We demonstrate that the accuracy can be improved by at least a factor of two over traditional precoders, and show that there is an optimal number of distinct precoders beyond which the CRB does not improve. 
	\end{abstract}
	
	\section{Introduction}
	
	\IEEEPARstart{M}{illimeter-wave} (mmWave) communication is expected to be one of the key enablers of 5th generation cellular networks \cite{roh2014millimeter}. Operating in mmWave frequencies offers a few advantages. First, there are large portions of underutilized bandwidth which could be used for multi-gigabit communications \cite{wang2015multi}. Second, due to the much shorter wavelength, MIMO systems consisting of many antennas can be compacted into much smaller sizes.
	However, in order to compensate for the stringent path loss characteristic of millimeter wave, highly directional beamforming is necessary at the transmitter and/or receiver \cite{kutty2016beamforming}. Since optimal precoding in communications can only be achieved after learning the channel, it is critical that fast and precise channel estimation techniques are developed. 
	
	The mmWave channel can be considered parsimonious in the sense that only a few multipath components carry non-negligible energy  \cite{rappaport2015wideband,keusgen2014propagation}. This structure inherent to the mmWave channel is leveraged in many techniques for performing quicker and/or finer channel estimation \cite{alkhateeb2014channel,bogale2015hybrid,xiao2016hierarchical,venugopal2016channel,rodriguez2016channel}.
	Channel estimation can be categorized in (a) initial access, or (b) tracking.
	Typically, the output of initial access is a set of angle-of-departures (AoD), angle-of-arrivals (AoA) and channel gains of the individual paths. 
	Due to the mobility of the users and the variability of the environment \cite{va2016millimeter,choi2016millimeter}, the validity of the initial access estimates come with an expiration time, and so it is necessary to estimate/track the channel periodically. The main idea behind tracking is that CSI is carried over to the next iteration, thus increasing the channel estimation accuracy and/or reducing the channel estimation overhead \cite{sanchis2002novel,seo2015training,he2014millimeter,kokshoorn2015fast}. 
	Because the channel parameters can be connected to the environment (such as is the case of geometric models), CSI may also be obtained through extraneous positioning and sensing technologies such as GPS, radars and cameras \cite{ditaranto2014location,choi2016millimeter}. For instance, if prior knowledge on the user position is available through GPS, then the set of possible directions between the base station and the user can be considerably reduced \cite{garcia2016location,aviles2016position,maschietti2017robust}.
	
	Whether initial access or tracking, the most common procedure for estimating the mmWave channel consists in sweeping the channel with beams at the transmitter and/or receiver \cite{nitsche2014ieee,wang2009beamforming,xiao2016hierarchical}. By detecting the time at which the received power is the largest, the correct pair of beams can be identified and the AoD and/or AoA estimated for each path. While intuitive, such beams may not necessarily yield the best achievable estimation accuracy. To the best of our knowledge, the literature in mmWave precoding has not addressed what are the fundamental limits in terms of AoA and AoD estimation when performing \emph{optimal precoding} at the transmitter.
	
	In this work, we seek to find the best transmit precoders for estimating the AoD and AoA, for a single-path channel, assuming that the AoD and AoA are known to lie within certain ranges of angles. Coarse knowledge on the AoD/AoA (CSI) may be available, for instance, through initial access or by tracking the AoD/AoA.
	To abstract the analysis from specific estimators, we use the Cram\'{e}r-Rao bound (CRB) from \cite{shahmansoori5g} as a proxy metric for the variance of the AoD and AoA. Indeed, the CRB is a lower bound on the variance of any unbiased estimator and it is tight at high SNR under some mild conditions \cite{van2004detection}. Our main contributions are:
	\begin{enumerate}
		\item Novel formulation of the optimal precoders according to the CRB on the AoD/AoA, with the option to include array gain constraints.
		\item Global optimization of the proposed non-convex problems by leveraging tools of convex optimization and majorization theory.
		\item Qualitative interpretation of the CRB expressions, and analysis of the optimal beampatterns.
	\end{enumerate}
	
	The remainder of this paper is organized as follows. In Section \ref{sec_signal_model} the system model and problem are described. Section \ref{sec:precoders} contains the formulation of the proposed precoders as well as the solution strategy and different extensions. Numerical results are presented in Section \ref{sec:results}, followed by conclusions in Section \ref{sec:conclusions}.

	\section{System Model} \label{sec_signal_model}
	
	Assume a transmitter (Tx) and a receiver (Rx) with $N_\mathrm{Tx}$ and $N_\mathrm{Rx}$ antennas, respectively. 
	The Tx sends $M$ consecutive training sequences (or pilots) consisting of $K$ symbols each, precoded by the vectors $\mathbf{f}_1,\ldots,\mathbf{f}_M\in\mathbb{C}^{N_\mathrm{Tx}\times 1}$.
	The precoders are normalized to $\|\mathbf{f}_m\|_2^2=\frac{1}{M}$ for all $m$, so that increasing its number does not result in increased transmitted energy.
	We assume a narrow-band model where the signal bandwidth is much smaller than the carrier frequency.
	The equivalent discrete baseband signal for the $k$-th symbol of the $m$-th training sequence at the Rx is
		\begin{equation} \label{eq:received_signal_full}
		\mathbf{y}_{m,k} = \mathbf{W}^\mathrm{H} \sum_{p=1}^{P}\alpha_p \,\mathbf{a}_\mathrm{Rx}(\phi_p) \mathbf{a}_\mathrm{Tx}^\mathrm{H}(\theta_p) \, \mathbf{f}_m s_k  + \mathbf{n}_{m,k},
		\end{equation}
		where $\alpha_p$, $\phi_p$, $\theta_p$ are the channel gain, AoA and AoD, respectively, of path $p$, $\mathbf{a}_\mathrm{Tx}(\theta)\in\mathbb{C}^{N_\mathrm{Tx}\times 1}$ and $\mathbf{a}_\mathrm{Rx}(\phi)\in\mathbb{C}^{N_\mathrm{Rx}\times 1}$ are the array responses at the Tx and Rx,
		$\mathbf{n}_{m,k}\sim\mathcal{CN}(\mathbf{0},\sigma^2\mathbf{I})$ is white\footnote{The noise is uncorrelated only if the $L$ combiners are pair-wise orthogonal, i.e., $\mathbf{W}^\mathrm{H}\mathbf{W}=\mathbf{I}$. An example of hybrid architecture with pair-wise orthogonal combiners is the case of arrays of subarrays where each RF chain is routed to a disjoint subset of antennas \cite{mendez2016hybrid}. If the combiners were not pair-wise orthogonal, by whitening the received signal through $(\mathbf{W}^\mathrm{H}\mathbf{W})^{-\frac{1}{2}}$, we would reach the same expression \eqref{eq:received_signal_after_combining} but with a different combining matrix $\widetilde{\mathbf{W}}=\mathbf{W}(\mathbf{W}^\mathrm{H}\mathbf{W})^{-\frac{1}{2}}$.} Gaussian noise, and $\mathbf{W}=[\mathbf{w}_1,\ldots,\mathbf{w}_L]\in\mathbb{C}^{N_\mathrm{Rx}\times L}$ is a fixed combining matrix. 
		The matrix  $\mathbf{W}$ may correspond to a hybrid array with $L<N_\mathrm{Rx}$ RF chains, e.g., when the Rx is a low-complexity user terminal. For the Tx we assume a fully digital architecture. 
		
		We assume that the $P$ paths are well-separated in terms of AoD and that the Tx knows that the dominant path ($p=1$) has an AoD and AoA which is known to be within a small range of angles\footnote{If the prior distributions of the AoD/AoA are provided instead, these ranges may be obtained from the confidence intervals.}:  $\theta_1\in\mathcal{R}_\mathrm{Tx}$ and $\phi_1\in\mathcal{R}_\mathrm{Rx}$. These ranges represent a priori uncertainty regarding the AoD/AoA, originating from mobility of the Rx, or from uncertainty in the tracking algorithm, or from uncertainty in location-aided communications \cite{garcia2016location,aviles2016position,maschietti2017robust}. This assumption allows us to simplify \eqref{eq:received_signal_full} to 
		\begin{equation} \label{eq:received_signal_after_combining}
		{\mathbf{y}}_{m,k} = \alpha_1 \, \mathbf{W}^\mathrm{H} \mathbf{a}_\mathrm{Rx}\left(\phi_1\right) \mathbf{a}_\mathrm{Tx}^\mathrm{H}\left(\theta_1\right) \mathbf{f}_m s_k  + {\mathbf{n}}_{m,k},
		\end{equation}
		provided $\mathbf{a}_\mathrm{Tx}^\mathrm{H}\left(\theta_k\right)\mathbf{f}_m \approx 0$ for $m \neq 1$. From now onward, we will omit the path index. Let $x^*$ be the complex conjugate of $x$. After coherent combining across time $k$ per precoder $\mathbf{f}_m$, we find $\mathbf{y}_m  = \sum_{k=1}^{K} s_k^* \, {\mathbf{y}}_{m,k}$, leading to the following matrix observation
		\begin{equation} \label{eq:received_matrix}
		\begin{split}
			\mathbf{Y} & = \frac{1}{\|\mathbf{s}\|_2}
			\begin{bmatrix}
			\mathbf{y}_1 & \cdots & \mathbf{y}_M
			\end{bmatrix} \\
			& =
			\alpha
			\|\mathbf{s}\|_2
			\mathbf{W}^\mathrm{H} \mathbf{a}_\mathrm{Rx}\left(\phi\right)\mathbf{a}_\mathrm{Tx}^\mathrm{H}\left(\theta\right) \mathbf{F}
			+\mathbf{N}
		\end{split}
		\end{equation}
		where $\mathbf{F} = [\mathbf{f}_1 \cdots \mathbf{f}_M]$, $\mathbf{s}=[s_1,\ldots,s_K]$,  and the components of $\mathbf{N}$ are i.i.d.\ Gaussian variables with variance $\sigma^2$. For notational convenience, and since we assume $\mathbf{W}$ to be fixed, we introduce $\mathbf{b}_\mathrm{Rx}\left(\phi\right)=\mathbf{W}^\mathrm{H} \mathbf{a}_\mathrm{Rx}\left(\phi\right)$. 
		
		Our goal is to find optimal precoders $\mathbf{F}$ to maximize the quality of the AoD and AoA estimates, given no knowledge of the channel gain $\alpha$. We tackle this problem by 
		minimizing the  CRB\footnote{The use of the CRB requires a high SNR operating condition. While mmWave communication operates under low SNR without beamforming \cite{andrews2017modeling}, our tracking scenario is congruent with a medium-to-high SNR assumption. } on the AoD $\theta$ and/or AoA $\phi$.

	\section{Optimal Precoders} \label{sec:precoders}
	
	To design the precoders, we choose as metric the CRB, which is a lower bound on the variance of any unbiased estimator. Such bound is well suited to this problem because it generally leads to tractable mathematical expressions, and more importantly, for a sufficiently large SNR and under some mild conditions \cite{van2004detection}, the variance of the maximum likelihood estimator (MLE) is tight to the CRB. If $\theta$ and $\phi$ are the AoD and AoA, respectively, of the LOS path, then determining $\theta$ yields the direction from the Tx to the Rx, and determining $\phi$ provides the Rx's orientation with respect to the Tx. Thus, for naming purposes and without loss of generality, we refer to the CRB on the $\theta$ as direction error bound (DEB), and the CRB on $\phi$ as orientation error bound (OEB). 
	Once the DEB and OEB are derived, we will proceed with the optimization of the precoder. 
	
	\subsection{Problem Formulation}
	
	From Appendix~\ref{sec:CRB}, the direction error bound is
	\begin{multline} \label{eq:bound:AoD}
	\operatorname{var}(\hat{\theta}) \geq
	\mathrm{DEB} =
	\Bigg[2 \, \mathrm{SNR}
	\left\|\mathbf{b}_\mathrm{Rx}\left(\phi\right)\right\|_2^2 \\
	\left(
	\left\|\mathbf{F}^\mathrm{H} \dot{\mathbf{a}}_\mathrm{Tx}\left(\theta\right)\right\|_2^2
	-\frac{\left|\mathbf{a}_\mathrm{Tx}^\mathrm{H} \left(\theta\right) \mathbf{F}  \mathbf{F}^\mathrm{H} \dot{\mathbf{a}}_\mathrm{Tx}\left(\theta\right)\right|^2}
	{\left\|\mathbf{F}^\mathrm{H} \mathbf{a}_\mathrm{Tx}\left(\theta\right)\right\|_2^2}
	\right)\Bigg]^{-1},
	\end{multline}
	and the orientation error bound is
	\begin{multline} \label{eq:bound:AoA}
	\operatorname{var}(\hat{\phi}) \geq
	\mathrm{OEB} =
	\Bigg[2 \, \mathrm{SNR}
	\left\|\mathbf{F}^\mathrm{H} \mathbf{a}_\mathrm{Tx}\left(\theta\right)\right\|_2^2 \\
	\left( \left\|\dot{\mathbf{b}}_\mathrm{Rx}\left(\phi\right)\right\|_2^2
	-\frac{\left|\mathbf{b}_\mathrm{Rx}^\mathrm{H}\left(\phi\right)´\dot{\mathbf{b}}_\mathrm{Rx}\left(\phi\right)\right|^2}
	{\left\|\mathbf{b}_\mathrm{Rx}\left(\phi\right)\right\|_2^2}
	\right)\Bigg]^{-1},
	\end{multline}
	where $\dot{\mathbf{a}}_\mathrm{Tx}(\theta) \triangleq
	\frac{\mathrm{d} \mathbf{a}_\mathrm{Tx}\left(\theta\right)}{\mathrm{d}\theta}$, $\dot{\mathbf{b}}_\mathrm{Rx}(\phi) \triangleq
	\frac{\mathrm{d} \mathbf{b}_\mathrm{Rx}\left(\phi\right)}{\mathrm{d}\phi}$ and
	\begin{equation} \label{eq:SNR}
	\mathrm{SNR} \triangleq \frac{|\alpha|^2 \|\mathbf{s}\|_2^2}{\sigma^2}.	
	\end{equation}
	For intuitive interpretations of the DEB and OEB see Appendix~\ref{sec:interpretation}.

	Both lower bounds depend on the precoders $\mathbf{F}$, but also on the AoD $\theta$ and AoA $\phi$ which are unknown. Next, we propose a min-max approach
	to the precoders design problem:
	\begin{problem} AoD-AoA-optimal precoders. Find the precoders that minimize the worst case DEB or OEB for all possible values of the AoD and AoA. \label{pro:problem:AODnAOA} 
		\begin{equation} \label{opt:problem:AODnAOA} 
		\min_{\substack{\left\|\mathbf{f}_m\right\|_2 = M^{-1} \\ m=1,\ldots,M}} \;
		\max_{(\theta,\phi)\in\mathcal{R}_{\mathrm{Tx}}\times\mathcal{R}_{\mathrm{Rx}}}
		\max \left\{\mathrm{DEB}, \mathrm{OEB}	\right\}
		\end{equation}
	\end{problem}
	The dependence of the DEB and OEB on $\theta$, $\phi$ and $\mathbf{F}$ has been omitted for notation clarity. Alternatively, if the only parameter of interest is the AoD, the optimal precoders are the solution to:
	\begin{problem} AoD-optimal precoders. Find the precoders that minimize the worst case DEB for all possible values of the AoD and AoA. \label{pro:problem:AOD} 
		\begin{equation} \label{opt:problem:AOD}
		\min_{\substack{\left\|\mathbf{f}_m\right\|_2 = M^{-1} \\ m=1,\ldots,M}} \;
		\max_{(\theta,\phi)\in\mathcal{R}_{\mathrm{Tx}}\times\mathcal{R}_{\mathrm{Rx}}}
		\mathrm{DEB}
		\end{equation}
	\end{problem}
	Conversely, if the only parameter of interest is the AoA, the optimal precoders are obtained from:
	\begin{problem} AoA-optimal precoders. Find the precoders that minimize the worst case OEB for all possible values of the AoD and AoA. \label{pro:problem:AOA} 
		\begin{equation} \label{opt:problem:AOA} 
		\min_{\substack{\left\|\mathbf{f}_m\right\|_2 = M^{-1} \\ m=1,\ldots,M}} \;
		\max_{(\theta,\phi)\in\mathcal{R}_{\mathrm{Tx}}\times\mathcal{R}_{\mathrm{Rx}}}
		\mathrm{OEB}
		\end{equation}
	\end{problem}

	\subsection{Convex Reformulation}
	
	Problems~\ref{pro:problem:AODnAOA}--\ref{pro:problem:AOA} are non-convex with respect to $\mathbf{F}$, and therefore, computing their global minimum efficiently is challenging. For a proof of non-convexity, note that the OEB is concave because it is proportional to $\|\mathbf{F}^\mathrm{H} \mathbf{a}_\mathrm{Tx}(\theta)\|^{-2}$, and the DEB depends on the  concave term $\|\mathbf{F}^\mathrm{H} \dot{\mathbf{a}}_\mathrm{Tx}(\theta)\|^{-2}$.
	In this section, Problems~\ref{pro:problem:AODnAOA}--\ref{pro:problem:AOA} will be reformulated as conic optimization problems, which is a subclass of convex problems. The conditions upon which the original problems and their convex reformulations are equivalent (in the sense that yield the same solution) will be analyzed in the next section.
	
	The analysis focuses in Problem~\ref{pro:problem:AODnAOA} because the solutions to Problems \ref{pro:problem:AOD} and \ref{pro:problem:AOA} will be shown to be subcases.
	An equivalent optimization problem to Problem~\ref{pro:problem:AODnAOA} is
	\begin{equation} \label{opt:max_min}
	\max_{\substack{\left\|\mathbf{f}_m\right\|_2 = M^{-1} \\ m=1,\ldots,M}} \;
	\min_{(\theta,\phi)\in\mathcal{R}_{\mathrm{Tx}}\times\mathcal{R}_{\mathrm{Rx}}}
	\min \left\{\mathrm{DEB}^{-1}, \mathrm{OEB}^{-1}	\right\}.
	\end{equation}
	By introducing a slack variable $t$, the above problem can be expressed in the hypograph form:
	\begin{subequations} \label{opt:max_min_wT}
		\begin{align}
		\max_{\mathbf{F},t} \quad& t \\
		\text{s.t.} \quad
		& \min_{(\theta,\phi)\in\mathcal{R}_{\mathrm{Tx}}\times\mathcal{R}_{\mathrm{Rx}}} \mathrm{DEB}^{-1} \geq t \label{opt:max_min_wT:DEB} \\
		& \min_{(\theta,\phi)\in\mathcal{R}_{\mathrm{Tx}}\times\mathcal{R}_{\mathrm{Rx}}} \mathrm{OEB}^{-1} \geq t \label{opt:max_min_wT:OEB} \\
		&\left\|\mathbf{f}_m\right\|_2 = \frac{1}{M} \qquad m=1,\ldots,M \label{opt:max_min_wT:energy}.
		\end{align}
	\end{subequations}

	\subsubsection*{Grid Approximation}
	
	The continuous set $\mathcal{R}_{\mathrm{Tx}}\times\mathcal{R}_{\mathrm{Rx}}$ makes optimizing over constraints \eqref{opt:max_min_wT:DEB}--\eqref{opt:max_min_wT:OEB} challenging. Instead we approximate it by a grid $\widetilde{\mathcal{R}}_{\mathrm{Tx}}\times\widetilde{\mathcal{R}}_{\mathrm{Rx}}$ such that
	\begin{align}
	\widetilde{\mathcal{R}}_{\mathrm{Tx}} &= 
	\left\{ \vartheta^1,\ldots,\vartheta^{S_\mathrm{Tx}}\right\} \subset \mathcal{R}_{\mathrm{Tx}} \label{eq:grid:AoD} \\
	\widetilde{\mathcal{R}}_{\mathrm{Rx}} &= 
	\left\{ \varphi^1,\ldots,\varphi^{S_\mathrm{Rx}}\right\} \subset \mathcal{R}_{\mathrm{Rx}}, \label{eq:grid:AoA}
	\end{align}
	where $S_{\mathrm{Tx}}$ and $S_{\mathrm{Rx}}$ are the number of discrete angles at the Tx and Rx, respectively, within the prior ranges.
	Also for notation brevity, define
	\begin{equation} \label{eq:short_notation}
	\begin{aligned}
	\mathbf{a}_\mathrm{Tx}^i &\triangleq \mathbf{a}_\mathrm{Tx}(\vartheta^i) & & & & &
	\mathbf{b}_\mathrm{Rx}^q &\triangleq \mathbf{b}_\mathrm{Rx}(\varphi^q) \\
	\dot{\mathbf{a}}_\mathrm{Tx}^i &\triangleq \dot{\mathbf{a}}_\mathrm{Tx}(\vartheta^i) & & & & &
	\dot{\mathbf{b}}_\mathrm{Rx}^q &\triangleq \dot{\mathbf{b}}_\mathrm{Rx}(\varphi^q).
	\end{aligned}
	\end{equation}
	Then, by replacing the continuous set by the grid, and since the DEB/OEB formulas \eqref{eq:bound:AoD}--\eqref{eq:bound:AoA} decouple into a part that depends only on $\theta$ and another on $\phi$, the left side of \eqref{opt:max_min_wT:DEB}--\eqref{opt:max_min_wT:OEB} can be expressed as
	\begin{align}
	\min_{(\theta,\phi)\in\mathcal{R}_{\mathrm{Tx}}\times\mathcal{R}_{\mathrm{Rx}}} &\mathrm{DEB}^{-1} =
	\nonumber \\ = 2 \, \mathrm{SNR} \, K_\text{D} & \min_{i\in\{1,\ldots,S_\mathrm{Tx}\}}\left(
	\left\|\mathbf{F}^\mathrm{H} \dot{\mathbf{a}}_\mathrm{Tx}^{i}\right\|_2^2
	-\frac{\left|\mathbf{a}_\mathrm{Tx}^{i \mathrm{H}} \mathbf{F} \mathbf{F}^\mathrm{H} \dot{\mathbf{a}}_\mathrm{Tx}^{i}\right|^2}
	{\left\|\mathbf{F}^\mathrm{H} \mathbf{a}_\mathrm{Tx}^{i}\right\|_2^2}\right) \label{eq:min_DEB} \\
	\min_{(\theta,\phi)\in\mathcal{R}_{\mathrm{Tx}}\times\mathcal{R}_{\mathrm{Rx}}} &\mathrm{OEB}^{-1} = 2 \, \mathrm{SNR} \, K_\text{O} \,
	\min_{i\in\{1,\ldots,S_\mathrm{Tx}\}}
	\left\|\mathbf{F}^\mathrm{H} \mathbf{a}_\mathrm{Tx}^{i}\right\|_2^2 \label{eq:min_OEB}
	\end{align}
	where $K_\text{D}$ and $K_\text{O}$ are the following constants (they do not depend on the optimizing variables $\mathbf{F}$ or $t$)
	\begin{align}
	K_\text{D} &\triangleq \min_{q\in \{1,\ldots,S_\mathrm{Rx}\}} \left\|\mathbf{b}_\mathrm{Rx}^q\right\|_{2}^{2} \label{eq:constant_DEB} \\
	K_\text{O} &\triangleq \min_{q\in \{1,\ldots,S_\mathrm{Rx}\}} \left(\left\|\dot{\mathbf{b}}_\mathrm{Rx}^q\right\|_2^2
	-\left|\mathbf{b}_\mathrm{Rx}^{q \mathrm{H}} \dot{\mathbf{b}}_\mathrm{Rx}^q\right|^2
	\left\|\mathbf{b}_\mathrm{Rx}^q\right\|_2^{-2}
	\right). \label{eq:constant_OEB}
	\end{align}
	Combining \eqref{opt:max_min_wT} with \eqref{eq:min_DEB}--\eqref{eq:min_OEB}, results in
	\begin{subequations} \label{opt:max_min_expand}
		\begin{align}
		\max_{\mathbf{F},t} \quad& t \\
		\text{s.t.} \quad
		& K_\text{D} \left(
		\left\|\mathbf{F}^\mathrm{H} \dot{\mathbf{a}}_\mathrm{Tx}^{i}\right\|_2^2
		-\frac{\left|\mathbf{a}_\mathrm{Tx}^{i \mathrm{H}} \mathbf{F} \mathbf{F}^\mathrm{H} \dot{\mathbf{a}}_\mathrm{Tx}^{i}\right|^2}
		{\left\|\mathbf{F}^\mathrm{H} \mathbf{a}_\mathrm{Tx}^{i}\right\|_2^2}\right) \geq t \label{opt:max_min_expand:DEB} \\
		& K_\text{O} \left\|\mathbf{F}^\mathrm{H} \mathbf{a}_\mathrm{Tx}^{i}\right\|_2^2 \geq t \label{opt:max_min_expand:OEB} \\
		&\left\|\mathbf{f}_m\right\|_2 = \frac{1}{M}  \qquad m=1,\ldots,M \label{opt:max_min_expand:energy}.
		\end{align}
	\end{subequations}
	for $i=1,\ldots,S_\mathrm{Tx}$.

	\subsubsection*{Relaxation on the Energy Constraint}
	
	In order to transform \eqref{opt:max_min_expand} into a convex problem, we propose a relaxation of the feasible set. If the solution to the newly relaxed problem is still within the original feasible set, then it is the optimum solution of the original problem. 
	The first proposed relaxation consists of replacing $\|\mathbf{f}_m\|_2 = M^{-1}$ for $m=1,\ldots,M$, by $\sum_{m=1}^{M} \|\mathbf{f}_m\|_2^2 = 1$. By taking into account that $\mathbf{f}_m$ is the $m$-th column of $\mathbf{F}$ and that $\|\mathbf{F}\|_\mathrm{F}^2=\trace \mathbf{F} \mathbf{F}^\mathrm{H}$, the new relaxed constraint can be expressed as
	\begin{equation}
	\trace\mathbf{F}\mathbf{F}^\mathrm{H}=1.
	\end{equation}

	\subsubsection*{Variable Change}
	
	Next, we introduce the following variable change
	\begin{equation} \label{eq:variable_change}
	\mathbf{X} = \mathbf{F}\mathbf{F}^\mathrm{H}, \quad\rank \mathbf{X} \leq M, \quad \mathbf{X}\succcurlyeq 0,
	\end{equation}
	where `$\succcurlyeq 0$' denotes positive semidefinite matrix. The variable change is reversible because  $\mathbf{F}$ can always be retrieved from $\mathbf{X}$ by, for instance, a Cholesky decomposition.

	\subsubsection*{Rank Relaxation}
	
	The second and final relaxation consists in dropping the rank constraint. Thus, the new optimization problem after including the two relaxations and performing the variable change is
	\begin{subequations} \label{opt:relaxed}
		\begin{align}
		\max_{\mathbf{X},t} \quad& t \\
		\text{s.t.} \quad
		& K_\text{D} \left(
		\dot{\mathbf{a}}_\mathrm{Tx}^{i\mathrm{H}} \mathbf{X} \dot{\mathbf{a}}_\mathrm{Tx}^{i}
		-\frac{\left|\mathbf{a}_\mathrm{Tx}^{i \mathrm{H}} \mathbf{X} \dot{\mathbf{a}}_\mathrm{Tx}^{i}\right|^2}
		{\mathbf{a}_\mathrm{Tx}^{i \mathrm{H}} \mathbf{X} \mathbf{a}_\mathrm{Tx}^{i}}\right) \geq t \label{opt:relaxed:DEB} \\
		& K_\text{O} \, \mathbf{a}_\mathrm{Tx}^{i \mathrm{H}} \mathbf{X} \mathbf{a}_\mathrm{Tx}^{i} \geq t \label{opt:relaxed:OEB} \\
		&\trace \mathbf{X} = 1 \label{opt:relaxed:energy} \\
		&\mathbf{X}\succcurlyeq 0
		\end{align}
	\end{subequations}
	for $i=1,\ldots,S_\mathrm{Tx}$.
	Constraint \eqref{opt:relaxed:DEB} can be cast as a second order cone \eqref{opt:conic:DEB} \cite[Chapter 2.3]{lobo1998applications},
	\begin{subequations} \label{opt:conic} 
		\begin{align}
		\max_{\mathbf{X},t} \quad
		&t \\
		&\left\|\begin{pmatrix}
		2 \mathbf{a}_\mathrm{Tx}^{i \mathrm{H}} \mathbf{X}\, \dot{\mathbf{a}}_\mathrm{Tx}^i \\
		\dot{\mathbf{a}}_\mathrm{Tx}^{i \mathrm{H}} \mathbf{X}\, \dot{\mathbf{a}}_\mathrm{Tx}^i
		-\frac{t}{K_\text{D}}
		-\mathbf{a}_\mathrm{Tx}^{i \mathrm{H}} \mathbf{X}\, \mathbf{a}_\mathrm{Tx}^i
		\end{pmatrix}\right\|_2 \nonumber\\
		&\qquad\qquad\leq \dot{\mathbf{a}}_\mathrm{Tx}^{i \mathrm{H}} \mathbf{X}\, \dot{\mathbf{a}}_\mathrm{Tx}^i
		-\frac{t}{K_\text{D}}
		+\mathbf{a}_\mathrm{Tx}^{i \mathrm{H}} \mathbf{X}\, \mathbf{a}_\mathrm{Tx}^i  \label{opt:conic:DEB} \\
		& K_\text{O} \, \mathbf{a}_\mathrm{Tx}^{i \mathrm{H}} \mathbf{X} \mathbf{a}_\mathrm{Tx}^{i} \geq t \label{opt:conic:OEB} \\
		&\trace \mathbf{X} = 1 \label{opt:conic:energy} \\
		&\mathbf{X}\succcurlyeq 0, \label{opt:conic:SD}
		\end{align}
	\end{subequations}
	where $\mathbf{X}\in\mathbb{C}^{N_\mathrm{Tx} \times N_\mathrm{Tx}}$ and $i=1,\ldots,S_\mathrm{Tx}$.
	Problem~\eqref{opt:conic} is a conic program \cite{boyd2004convex}, and consequently convex, because it is composed of linear constraints, second order cones \eqref{opt:conic:DEB} and a positive semidefinite cone \eqref{opt:conic:SD}. An advantage of having reformulated our problem as a conic program is that they are well studied in the literature and very efficient solvers exist \cite{grant2008cvx,Mosek2016}.
	
	Solving Problem~\ref{pro:problem:AODnAOA} requires knowledge of the combining matrix $\mathbf{W}$ because it appears in constants $K_\text{D}$ \eqref{eq:constant_DEB} and $K_\text{D}$ \eqref{eq:constant_OEB} in constraints \eqref{opt:conic:DEB} and \eqref{opt:conic:DEB}, respectively. If the combining matrix of choice $\mathbf{W}$ is more beneficial for AoA estimation than AoD estimation, then the resulting constants, $K_\text{D}$  and $K_\text{O}$, will balance it out by favoring the optimization of the DEB (which is a bound on the AoD accuracy) over the OEB. Vice-versa, if $\mathbf{W}$ is more beneficial for AoD estimation, then the resulting constants will favor the optimization of the OEB.
	For the solution to Problem~\ref{pro:problem:AOD} simply solve the same problem \eqref{opt:conic} without constraint \eqref{opt:conic:OEB}. By deleting \eqref{opt:conic:OEB} and performing the variable change $\frac{t}{K_\text{D}}\leftarrow t$, constants $K_\text{O}$ and $K_\text{D}$ disappear, making the optimization problem independent of the Rx's array response. To obtain the solution to Problem~\ref{pro:problem:AOA} delete \eqref{opt:conic:DEB}, and following the same reasoning, the optimization problem also becomes independent of the Rx's array response.

	\begin{figure}
		\centering
		\includegraphics[width=\columnwidth]{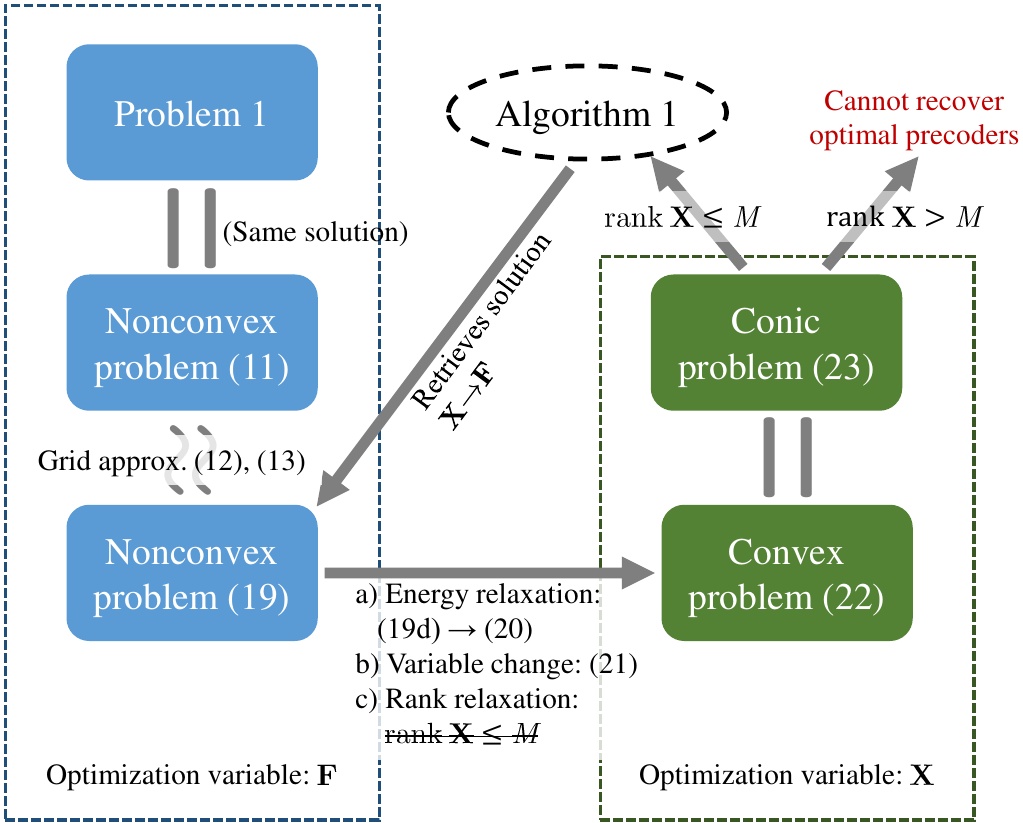}
		\caption{Flow chart of the multiple steps for obtaining a tractable optimization problem. The nonconvex and convex reformulations are colored in blue and green, respectively. The optimal precoders are obtained by optimizing problem~\eqref{opt:conic} and executing Algorithm~\ref{alg:optimal_precoders}. We have verified numerically that the grid approximation is accurate when sufficiently.}
		\label{fig:flow_chart}
	\end{figure}

	\subsection{Recovery of the Precoders} \label{sec:recovery_precoders}
	
	In the previous section, Problem~\ref{pro:problem:AODnAOA} was transformed into a convex problem~\eqref{opt:conic} by
	\begin{itemize}
		\item[A)] approximating the prior ranges with a grid \eqref{eq:grid:AoD}, \eqref{eq:grid:AoA},
		\item[B)] relaxing the feasible set by replacing $\|\mathbf{f}_m\|_2 = M^{-1}$ for $m=1,\ldots,M$, by $\sum_{m=1}^{M} \|\mathbf{f}_m\|_2^2 = 1$.
		\item[C)] change of variables $\mathbf{X} = \mathbf{F}\mathbf{F}^\mathrm{H}$, $\rank \mathbf{X} \leq M$, $\mathbf{X}\succcurlyeq 0$,
		\item[D)] relaxing the feasible set by dropping $\rank\mathbf{X}\leq M$.
	\end{itemize}
	We abuse the notation and call $\mathbf{X}$ the global optimum of \eqref{opt:conic} instead of the optimizing variable.
	It turns out that the grid approximation (A) is very accurate when made dense enough.
	Moreover, the trace constraint in \eqref{opt:conic:energy} induces a low-rank solution \cite{recht2010guaranteed,fazel2001rank}. Hence, if the number of precoders $M$ is sufficiently large, then it will occur that $\rank\mathbf{X}\leq M$, and the relaxation (D) will not affect the solution.
	Thus, in the remainder of this section we discuss the effect of (B) and (C)
	assuming (A) is exact and (D) is satisfied. These two later assumptions will be validated numerically in Section~\ref{numerical_results:optimality}.
	
	Given $\mathbf{X}$, the variable change (C) can be reversed by performing a Cholesky decomposition \cite{golub2012matrix}.
	However, we are interested in a decomposition from $\mathbf{X}$ to $\mathbf{F}$ that also satisfies $\|\mathbf{f}_m\|_2 = M^{-1}$ for $m=1,\ldots,M$ (B). If such decomposition existed, then $\mathbf{F}$ could be recovered from $\mathbf{X}$ and satisfy all the constraints to the original Problem~\ref{pro:problem:AODnAOA} in its hypograph form \eqref{opt:max_min_wT}. 
	In summary, we need to find $\mathbf{F}$ such that
	\begin{description}
		\item[C1)] $\mathbf{F}\mathbf{F}^\mathrm{H} = \mathbf{X}$
		\item[C2)] $\diag\left(\mathbf{F}^\mathrm{H}\mathbf{F}\right)= M^{-1}\mathbf{1}$
	\end{description}
	where $\diag(\mathbf{F}^\mathrm{H}\mathbf{F})$ denotes a vertical vector stacking the entries on the main diagonal of $\mathbf{F}^\mathrm{H}\mathbf{F}$.
	Define $R\triangleq\rank \mathbf{X}\leq M$ where the latter inequality follows from assumption (D). From C1 we infer that $\rank \mathbf{F} = R$.
	Sufficient conditions for satisfying C1 and C2 can be obtained by expanding $\mathbf{F}$ and $\mathbf{X}$ by their \emph{compact} singular value decomposition (SVD) and eigendecomposition (ED), respectively, 
	\begin{align}
	\mathbf{F} &= \mathbf{U}\mathbf{\Sigma}\mathbf{V}^\mathrm{H} \label{eq:SVD_F} \\
	\mathbf{X} &= \mathbf{Q}\mathbf{\Lambda}\mathbf{Q}^\mathrm{H} \label{eq:ED_X},
	\end{align}
	where $\mathbf{U}=[\mathbf{u}_1 \cdots \mathbf{u}_{R}]\in\mathbb{C}^{N_\mathrm{Tx}\times R}$ is a matrix of left singular vectors, $\mathbf{V}=[\mathbf{v}_1 \cdots \mathbf{v}_{R}]\in\mathbb{C}^{R\times R}$ is a matrix of right singular vectors, $\mathbf{Q}=[\mathbf{q}_1 \cdots \mathbf{q}_{R}]\in\mathbb{C}^{N_\mathrm{Tx}\times R}$ is a matrix of eigenvectors, and $\mathbf{\Sigma}$ and $\mathbf{\Lambda}$ are diagonal matrices whose diagonal elements are the singular values $\mu_{1}\cdots\mu_{R}$ and the eigenvalues $\lambda_{1}\cdots\lambda_{R}$, respectively. 
	
	Substituting $\mathbf{F}$ and $\mathbf{X}$ in C1 with \eqref{eq:SVD_F} and \eqref{eq:ED_X}, respectively, leads to a new form of C1, $\mathbf{U}\mathbf{\Sigma}^2\mathbf{U}^\mathrm{H} = \mathbf{Q}\mathbf{\Lambda}\mathbf{Q}^\mathrm{H}$.
	This equation is solved by matching the eigenvectors and eigenvalues on both sides of the equation, i.e.,
	\begin{align}
	\mathbf{u}_{m} &= \mathbf{q}_{m} \label{eq:F_leftSingVecs} \\
	\mu_{m} &= \sqrt{\lambda_{m}}, \label{eq:F_singVals}
	\end{align}
	for $m=1,\ldots,R$. Similarly, substituting  \eqref{eq:SVD_F} and \eqref{eq:ED_X} into C2 yields the new condition
	\begin{equation} \label{eq:condition_constant_energy}
	\diag\left(\mathbf{V}\mathbf{\Sigma}^{2}\mathbf{V}^\mathrm{H}\right) = M^{-1}\mathbf{1}.
	\end{equation}
	The (diagonal) elements of $\mathbf{\Sigma}$ have been fixed by \eqref{eq:F_singVals}, hence, only $\mathbf{V}$ can be adjusted to meet the new condition.
	Notice that $\mathbf{Z}\triangleq \mathbf{V}\Sigma^{2}\mathbf{V}^\mathrm{H}$ may be regarded as the eigendecomposition of a matrix whose non-zero eigenvalues are $\mu_1^2\ldots,\mu_R^2$ and their corresponding eigenvectors $\mathbf{v}_1,\ldots,\mathbf{v}_R$.
	Consequently, condition \eqref{eq:condition_constant_energy} is equivalent to the existence of a matrix $\mathbf{Z}$ with prescribed non-zero eigenvalues $\mu_1^2\ldots,\mu_R^2$ and $M^{-1}$ in all entries of the main diagonal.
	Fortunately, for the particular case of constant diagonal, the Schur-Horn theorem \cite[B.1.\ in p.~218]{olkin2016inequalities} states that $\mathbf{Z}$ always exists, and it can be obtained by the Bendel-Mickey algorithm \cite{bendel1978population,davies2000numerically}.
	
	In summary, the different pieces of a valid $\mathbf{F}$ satisfying C1 and C2 can be obtained as follows: its left singular vectors by \eqref{eq:F_leftSingVecs}, its singular values by \eqref{eq:F_singVals} and its right singular vectors by finding a matrix satisfying \eqref{eq:condition_constant_energy} with the Bendel-Mickey algorithm and computing its eigenvectors.
	All these steps are summarized in Algorithm~\ref{alg:optimal_precoders}.
	\begin{algorithm} \caption{Retrieval of the optimal precoders $\mathbf{F}$ from $\mathbf{X}$} \label{alg:optimal_precoders}
		\begin{algorithmic}[1]
			
			\Statex \textbf{inputs:} $\mathbf{X}, M$
			
			\Statex \textbf{outputs:} $\mathbf{F}$
			
			\State compute the \emph{compact} ED of $\mathbf{X} = \mathbf{Q} \mathbf{\Lambda} \mathbf{Q}^\mathrm{H}$ where the diagonal elements of $\mathbf{\Lambda}$ are $\lambda_1,\ldots,\lambda_R$ and $R=\rank \mathbf{X}$
			
			\State find a matrix $\mathbf{Z}$ with eigenvalues $\lambda_1,\ldots,\lambda_{R}$ and all diagonal elements equal to $M^{-1}$ with the Bendel-Mickey algorithm\footnotemark \cite{bendel1978population,davies2000numerically} \label{line:Bendel-Mickey}
			
			\State compute the ED of $\mathbf{Z}$ and denote $\mathbf{v}_1,\ldots,\mathbf{v}_R$ the eigenvectors associated to eigenvalues $\lambda_1,\ldots,\lambda_R$
			
			\State set $\mathbf{F}= \mathbf{Q} \mathbf{\Lambda}^{\frac{1}{2}} [\mathbf{v}_1 \cdots \mathbf{v}_R]^\mathrm{H}$
			
		\end{algorithmic}
	\end{algorithm}
	\footnotetext{Code available in Matlab R2015b by the name \texttt{gallery(\textquotesingle randcorr\textquotesingle,\emph{arg})}.}
	
	\paragraph*{Remark} The solution to \eqref{eq:condition_constant_energy} is in general not unique, and so the Bendel-Mickey algorithm is designed to return a random solution in the solution set. Consequently, the solution to Problems~\ref{pro:problem:AODnAOA}--\ref{pro:problem:AOA} is random as well.
	
	\subsubsection*{Examples}
	
	Define the array gain at the Tx as
	\begin{equation} \label{eq:array_gain}
		g(\theta,\mathbf{f}_m) \triangleq \left| \mathbf{f}_m^\mathrm{H} \mathbf{a}_\mathrm{Tx}(\theta)\right| / \left\|\mathbf{f}_m\right\|,
	\end{equation}
	and the aggregated gain  (which is the equivalent gain at the Rx if the energy of the $M$ training sequences is coherently combined) as
	\begin{equation} \label{eq:total_gain}
		g_\text{T}(\theta,\mathbf{F}) \triangleq \sqrt{\sum_{m=1}^{M} |g(\theta,\mathbf{f}_m)|^2} = \sqrt{M}
	\left\| \mathbf{F}^\mathrm{H} \mathbf{a}_\mathrm{Tx}(\theta)\right\|_2,
	\end{equation}
	where we used the fact that the norm of any precoder is $M^{-1}$.
	
	\begin{figure*}
		\centering
		\begin{tikzpicture}
		\begin{axis}[
		hide axis,
		width=1.7cm,
		cycle list name=myCycleList,
		legend style={at={(0,.6)},anchor=north west},
		legend cell align=left,
		]
		\addplot[white,forget plot]{0};
		\addlegendimage{no markers,lines-1,thick}
		\addlegendentry{Gain precoder $\mathbf{f}_1$}
		\addlegendimage{no markers,lines-2,thick}
		\addlegendentry{Gain precoder $\mathbf{f}_2$}
		\addlegendimage{no markers,lines-3,thick}
		\addlegendentry{Gain precoder $\mathbf{f}_3$}
		\addlegendimage{no markers,black,thick}
		\addlegendentry{Aggregated gain}
		\end{axis}
		\end{tikzpicture} \hfil
		\subfloat[AoD-AoA-optimal precoders (Problem~\ref{pro:problem:AODnAOA}). The rank of $\mathbf{X}$ is 2.]{
			\includegraphics[trim={2.2cm 0 2.2cm 0},clip,width=3.8cm]{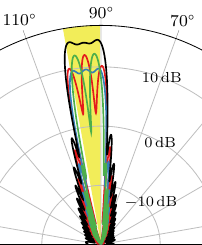}
			\label{fig:precoders:optimal} }
		\hfil
		\subfloat[AoD-optimal precoders (Problem~\ref{pro:problem:AOD}). The rank of $\mathbf{X}$ is 2.]{
			\includegraphics[trim={2.2cm 0 2.2cm 0},clip,width=3.8cm]{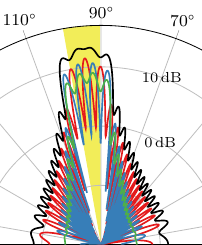}
			\label{fig:precoders:AoD-optimal} }
		\hfil
		\subfloat[AoA-optimal precoders (Problem~\ref{pro:problem:AOA}). The rank of $\mathbf{X}$ is 3.]{
			\includegraphics[trim={2.2cm 0 2.2cm 0},clip,width=3.8cm]{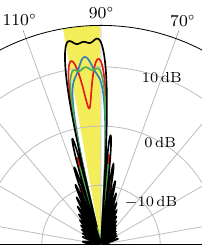}
			\label{fig:precoders:AoA-optimal} }
		\caption{Array gain of the precoders vs.\ azimuth for all $M$ precoders. The Tx and Rx are equipped with 30-antenna half-wavelength inter antenna spacing uniform linear array (ULA) \cite{tsai2004ber}. The number of training sequences is $M=3$. The AoD (same for the AoA for simplicity) is known a priori to lie in the range $[\SI{90}{\degree},\SI{100}{\degree}]$, indicated by the yellow shaded area. According to Section~\ref{sec:recovery_precoders}, the optimality of these precoders is ensured because for all cases $M\geq\rank\mathbf{X}$.}
		\label{fig:precoders}
	\end{figure*}
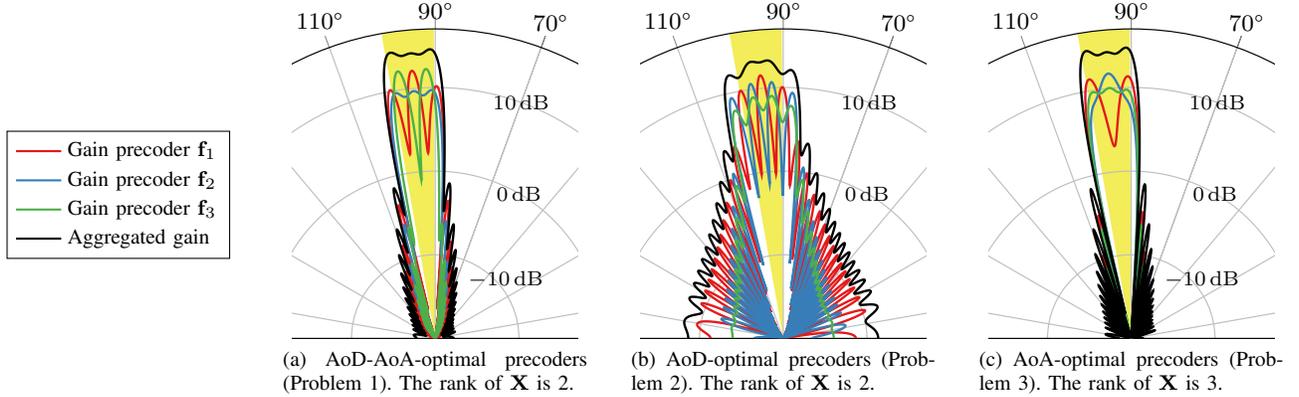
	
	Fig.~\ref{fig:precoders} plots the beampatterns (gain vs.\ azimuth) of the AoD-AoA-optimal precoders. The gain and aggregated gain\footnote{To convert the (aggregated) gain to \si{\decibel} we apply the function $20\log_{10}(\cdot)$.} are defined in \eqref{eq:array_gain} and \eqref{eq:total_gain}. The yellow zone in the figures represents the prior range on the AoD. Furthermore, we define the term `worst aggregated gain', as the lowest aggregated gain within the prior range.
	As expected from the interpretation in Section~\ref{sec:precoders},
	where it was argued that the OEB is optimized when the gain $\|\mathbf{F}^\mathrm{H} \mathbf{a}_\mathrm{Tx}(\theta)\|_2$ is maximized, in Fig.~\ref{fig:precoders:AoA-optimal} the AoA-optimal precoders' worst aggregated gain is \SI{14.1}{\decibel}, which is the highest among the three subfigures.
	The AoD-optimal precoders in Fig.~\ref{fig:precoders:AoD-optimal} show a large number of ripples, which is also consistent with the interpretation in Section~\ref{sec:precoders}. 
	The intuitive explanation is that the Rx identifies the AoD by observing the changes of received signal strength and phase changes for the $M$ training sequences. Thus, in order to increase the AoD estimation accuracy, the Tx's precoders gain and phase for closely spaced AoD must be as different as possible. The AoD-optimal precoders worst aggregated gain, \SI{13.6}{\decibel}, is just slightly worse than that of the AoA-optimal precoders.
	The AoD-AoA-optimal precoders offer a trade-off between the AoA-optimal and AoD-optimal precoders in terms of aggregated gain and `ripples'. 
	Because the AoA estimation accuracy is optimized by maximizing the aggregated gain towards the Rx, the AoD-AoA- and AoA-optimal precoders spill much less energy outside the prior Tx range of angles than the AoD-optimal precoders. To explicitly reduce energy transmitted outside the prior Tx range, Section~\ref{sec:attenuation}, introduces an extra constraint that when added to problem \eqref{opt:conic} enforces a certain out-of-range attenuation.

	\subsection{Additional Constraints}
	
	The proposed precoders are purely designed to minimize the CRB of the AoD and/or AoA. However, in practice some aditional constraints may be desired, such as null steering towards certain directions in order to mitigate multiuser interference or cancel certain paths. Next, it is shown how some of these constraints can easily be added to the original optimization problems.
	
	\subsubsection{Identifiability of the AoD and AoA} \label{sec:ambiguities}
	
	A necessary condition for correctly estimating the AoD and/or AoA is that they are identifiable. Otherwise, even in the absence of noise, we may suffer from large estimation errors due to signal ambiguities.
	The AoA and AoD are identifiable provided that the received signals \eqref{eq:received_matrix} in absence of noise are different
	\begin{equation}
	\alpha \, \mathbf{b}_\mathrm{Rx}(\phi) \mathbf{a}_{\mathrm{Tx}}^{\mathrm{H}}(\theta) \mathbf{F}
	\neq
	\alpha' \, \mathbf{b}_\mathrm{Rx}(\phi') \mathbf{a}_{\mathrm{Tx}}^{\mathrm{H}}(\theta') \mathbf{F},
	\end{equation}
	for all possible $\theta\neq\theta'$ in $\mathcal{R}_\mathrm{Tx}$, $\phi\neq\phi'$ in $\mathcal{R}_\mathrm{Rx}$, or $\alpha \neq \alpha'$. Since the signals are rank 1 matrices, by applying the substitution $\mathbf{b}_\mathrm{Rx}(\phi) =\mathbf{W}^\mathrm{H}\mathbf{a}_\mathrm{Rx}(\phi)$ the condition decouples into pair-wise linear independence between steering vectors at the Rx and Tx:
	\begin{align} \label{eq:identify:AoA}
	\mathbf{W}^\mathrm{H}\mathbf{a}_\mathrm{Rx}(\phi) &\neq \beta \, \mathbf{W}^\mathrm{H}\mathbf{a}_\mathrm{Rx}(\phi') \\
	\label{eq:identify:AoD}
	\mathbf{F}^{\mathrm{H}} \mathbf{a}_\mathrm{Tx}(\theta) &\neq \gamma \, \mathbf{F}^{\mathrm{H}}\mathbf{a}_\mathrm{Tx}(\theta')
	\end{align}
	for all complex $\beta$ and $\gamma$, different than zero.
	From these expressions we can conclude that the combining matrix impacts the identifiability of the AoA, and that the precoding matrix impacts the identifiability of the AoD only.
		For the particular case of a fully digital Rx ($\mathbf{W}=\mathbf{I}$), the AoA is identifiable for most common types of arrays \cite{godara1981uniqueness,gazzah2009optimum}, whereas for arbitrary combiners, the conditions for which the AoA is identifiable are not known in the literature to the best of the authors' knowledge (and also beyond the scope of the current study).

		Since this work deals with the design of the precoders, we turn to condition \eqref{eq:identify:AoD} which imposes pair-wise independence and can also be expressed as
		$|\mathbf{a}_{\mathrm{Tx}}^\mathrm{H} (\theta) \mathbf{F}\, \mathbf{F}^\mathrm{H}\, \mathbf{a}_{\mathrm{Tx}}(\theta')|^2
		\|\mathbf{F}^\mathrm{H}\,\mathbf{a}_{\mathrm{Tx}}(\theta)\|_2^{-2} \|\mathbf{F}^\mathrm{H}\,\mathbf{a}_{\mathrm{Tx}}(\theta')\|_2^{-2} < 1$.
	Thus, in order to ensure that the AoD is identifiable we propose to add the following constraints to problem \eqref{opt:conic}:
	\begin{equation} \label{eq:correlation}
	\frac{\left|\mathbf{a}_{\mathrm{Tx}}^\mathrm{H} (\vartheta^{i}) \mathbf{F}\, \mathbf{F}^\mathrm{H}\, \mathbf{a}_{\mathrm{Tx}}(\vartheta^{i'})\right|^2}
	{\left\|\mathbf{F}^\mathrm{H}\,\mathbf{a}_{\mathrm{Tx}}(\vartheta^{i})\right\|_2^2 \left\|\mathbf{F}^\mathrm{H}\,\mathbf{a}_{\mathrm{Tx}}(\vartheta^{i'})\right\|_2^2} < \rho
	\end{equation}
	for all $i,i'$ such that such that $\vartheta^{i}-\vartheta^{i'}>D (\operatorname{mod}2\pi)$,
	where $0\leq\rho<1$ and $D$ is the angular resolution of the array\footnote{The resolution of the array is typically related to the antenna aperture and is different for each array configuration. In practice, one could start with an approximate value and then tweak this parameter. For instance, for an $N$-antennas half-wavelength inter-antenna spaced uniform circular  array (UCA) \cite{tsai2004ber}, the angle resolution has been checked numerically to be well approximately by $1.6\sin(\pi/N)$ [radians].}. By performing the variable change $\mathbf{X}=\mathbf{F} \mathbf{F}^\mathrm{H}$ and some algebraic manipulations \cite{lobo1998applications}, \eqref{eq:correlation} can be expressed as a second order cone (convex constraint):
	\begin{multline} \label{eq:correlation_constr_convex}
	\left\|
	\begin{pmatrix}
	2 \, \mathbf{a}_{\mathrm{Tx}}^{i \mathrm{H}} \mathbf{X} \, \mathbf{a}_{\mathrm{Tx}}^{i'} \\
	\sqrt{\rho} \, \mathbf{a}_{\mathrm{Tx}}^{i \mathrm{H}} \mathbf{X} \, \mathbf{a}_{\mathrm{Tx}}^{i}
	- \sqrt{\rho} \, \mathbf{a}_{\mathrm{Tx}}^{i' \mathrm{H}} \mathbf{X} \, \mathbf{a}_{\mathrm{Tx}}^{i'}
	\end{pmatrix} \right\|_2 \\
	\leq \sqrt{\rho} \, \mathbf{a}_{\mathrm{Tx}}^{i \mathrm{H}} \mathbf{X} \, \mathbf{a}_{\mathrm{Tx}}^{i}
	+ \sqrt{\rho} \, \mathbf{a}_{\mathrm{Tx}}^{i' \mathrm{H}} \mathbf{X} \, \mathbf{a}_{\mathrm{Tx}}^{i'},
	\end{multline}
	where the shortened notation \eqref{eq:short_notation} was used. This condition must be added to optimization problem \eqref{opt:conic} whenever we wish to estimate the AoD because its identifiability.

	\subsubsection{Out-of-Range Attenuation} \label{sec:attenuation}
	
	From Fig.~\ref{fig:precoders}, we can observe that the precoders  radiate some non-negligible energy in directions outside the prior range $\mathcal{R}_\mathrm{Tx}$,
	even though an initial assumption was that the precoders do not illuminate any paths other than the desired one.
	In addition, there may be some operational constraints such as low sidelobe ratio \cite{venkateswaran2016hybrid} for improved inter-user interference or the placement of nulls in certain areas of the beampattern.
	
	Let $\{\pi^q\}_{q=1}^{S_\text{a}}$ be the AoDs for which we wish to enforce a lower transmitting power, and recall $\{\vartheta^i\}_{i=1}^{S_\mathrm{Tx}}$ is the grid of angles within the prior range \eqref{eq:grid:AoD}.
	The total transmitted energy in direction $\pi^q$ over the $M$ training sequences is $\|\mathbf{F}^\mathrm{H} \mathbf{a}_{\mathrm{Tx}}(\pi^q)\|_2^2$.
	Let $A_q<1$ be the desired attenuation factor, then, we propose to add the following constraint to problem \eqref{opt:conic},
	\begin{equation}
	\left\|
	\mathbf{F}^\mathrm{H} \mathbf{a}_{\mathrm{Tx}}\left(\pi^q\right)
	\right\|_2^2
	\leq
	A_q
	\left\|
	\mathbf{F}^\mathrm{H} \mathbf{a}_{\mathrm{Tx}}\left(\vartheta^{i}\right)
	\right\|_2^2
	\end{equation}
	for $q=1,\ldots,S_\text{a}$ and $i=1,\ldots,S_\mathrm{Tx}$, which can be transformed to linear constraints after performing the variable change $\mathbf{X}=\mathbf{F} \mathbf{F}^\mathrm{H}$,
	\begin{equation} \label{opt:Rx}
	\mathbf{a}_{\mathrm{Tx}}^\mathrm{H}\left(\pi^q\right)
	\mathbf{X}\,
	\mathbf{a}_{\mathrm{Tx}}\left(\pi^q\right)
	\leq
	A_q\,
	\mathbf{a}_{\mathrm{Tx}}^\mathrm{H} \left(\vartheta^{i}\right)
	\mathbf{X}\,
	\mathbf{a}_{\mathrm{Tx}}\left(\vartheta^{i}\right).
	\end{equation}
	These $S_\text{a}S_\mathrm{Tx}$ constraints can be reduced to $S_\text{a}+S_\mathrm{Tx}$ by incorporating a dummy variable $z$,
	\begin{alignat}{4}
	\mathbf{a}_{\mathrm{Tx}}^\mathrm{H}\left(\pi^q\right)
	\mathbf{X}\,
	\mathbf{a}_{\mathrm{Tx}}\left(\pi^q\right)
	&\leq A_q\,z
	&\qquad q&=1,\ldots,S_\text{a} \label{opt:multi-user} \\
	\mathbf{a}_{\mathrm{Tx}}^\mathrm{H} \left(\vartheta^{i}\right)
	\mathbf{X}\,
	\mathbf{a}_{\mathrm{Tx}}\left(\vartheta^{i}\right) &\geq z
	&\qquad i&=1,\ldots,S_\mathrm{Tx}. \label{opt:multi-user2}
	\end{alignat}

	\subsubsection*{Examples}
	
\begin{figure}
		\centering
		
		\begin{tikzpicture}
		\begin{axis}[
		hide axis,
		scale only axis,
		height=0pt,
		width=0pt,
		colorbar horizontal,
		point meta min=0,
		point meta max=1,
		colorbar style={
			width=.8\columnwidth,
			height=.5em,
			xtick={0,.2,.4,.6,.8,1}}
		]
		\addplot [draw=none] coordinates {(0,0)};
		\end{axis}
		\end{tikzpicture}  \\
		\hspace{.2cm}
		\subfloat[Without identifiability constraints \eqref{eq:correlation}.]{
			\includegraphics{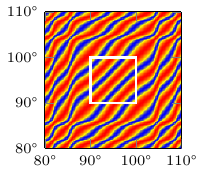}
			\label{fig:ambiguities:not_identifiable}
		}
		\hspace{.8cm}
		\subfloat[With identifiability constraints \eqref{eq:correlation} and $\rho=0.1$.]{
			\includegraphics{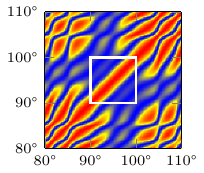}
			\label{fig:ambiguities:identifiable}
		}
		\caption{Colormap of left hand side of \eqref{eq:correlation} versus $\theta$ and $\theta'$ when employing AoD-AoA-optimal precoders. The Tx and the Rx have, each, a 30-antennas uniform circular array (UCA) \cite{tsai2004ber}. The number of training sequences $M=4$ and the rank of $\mathbf{X}$ is 4. It is known a priori that the AoD lies in the interal $\mathcal{R}_\mathrm{Tx}=[\SI{70}{\degree},\SI{90}{\degree}]$. The white square indicates the region $\mathcal{R}_\mathrm{Tx} \times \mathcal{R}_\mathrm{Tx}$.}
		\label{fig:ambiguities}
	\end{figure}
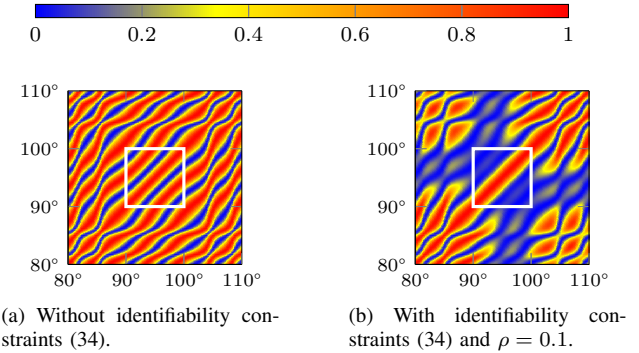

	Fig.~\ref{fig:ambiguities} plots the left hand side of \eqref{eq:correlation} when the precoders are obtained with or without the identifiability constraint. 
	Ideally, only the anti-diagonal across the white square should be red as is the case in the right figure. The left figure, has two red stripes, and therefore, there are pairs of AoDs within the range $\mathcal{R}_\mathrm{Tx}$ which result in the same signals, and consequently are not identifiable even in the absence of noise.

\begin{figure}
		\centering
		\includegraphics{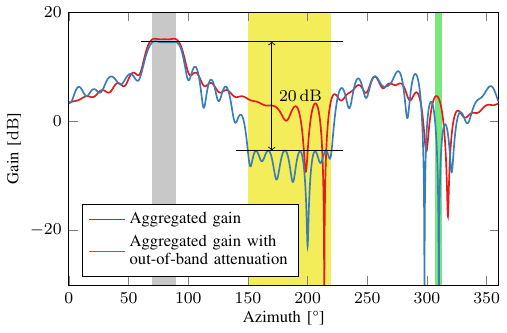}	
		\caption{Aggregated gain \eqref{eq:total_gain} of the AoD-AoA-optimal precoders vs.\ azimuth. The Tx and Rx are equipped with 30-antenna half-wavelength inter antenna spacing UCA. The number of training sequences is $M=3$ and the rank $\mathbf{X}$ is 2, ensuring the precoders are optimal. The AoD (same for the AoA for simplicity) is known a priori to lie in the range $[\SI{70}{\degree},\SI{90}{\degree}]$, indicated by the grey shaded bin. Two different sets of precoders are used. First set is the optimum solution to Problem~\ref{pro:problem:AODnAOA}. The second set is the optimum solution to Problem~\ref{pro:problem:AODnAOA} while imposing a \SI{20}{\decibel} attenuation in the range [\SI{150}{\degree},\SI{220}{\degree}] indicated by the yellow shaded bin, and a null at \SI{310}{\degree} indicated by the green shaded bin.	}
		\label{fig:gain}
	\end{figure}

	In Fig.~\ref{fig:gain} two sets of AoD-AoA-optimal precoders are generated. The first set of precoders has no constraint on the aggregated gain \eqref{eq:total_gain} towards directions outside of the range of interest $\mathcal{R}_\mathrm{Tx}$. The second set of precoders imposes a \SI{20}{\decibel} attenuation in the range [\SI{150}{\degree},\SI{220}{\degree}] and a null at \SI{310}{\degree}. Note that the aggregated gain within the range is virtually the same for both sets of precoders.

	\section{Numerical Results} \label{sec:results}
	
	The performance of the optimal precoders is illustrated next. To this end, we define the following figures of merit which are in fact the square roots of the objective functions of
	Problems~\ref{pro:problem:AODnAOA}--\ref{pro:problem:AOA}, respectively,
	\begin{align}
	\text{Worst case rEB} =& \max_{(\theta,\phi)\in\mathcal{R}_{\mathrm{Tx}}\times\mathcal{R}_{\mathrm{Rx}}} \sqrt{\mathrm{EB}}  \label{eq:metrics:AoD-AoA} \\
	\text{Worst case rDEB} =& \max_{(\theta,\phi)\in\mathcal{R}_{\mathrm{Tx}}\times\mathcal{R}_{\mathrm{Rx}}} \sqrt{\mathrm{DEB}}  \label{eq:metrics:AoD} \\
	\text{Worst case rOEB} =& \max_{(\theta,\phi)\in\mathcal{R}_{\mathrm{Tx}}\times\mathcal{R}_{\mathrm{Rx}}} \sqrt{\mathrm{OEB}}, \label{eq:metrics:AoA}
	\end{align}
	where $\mathrm{EB}=\max\{\mathrm{DEB}, \mathrm{OEB}\}$ and the `r' denotes square root.
		The following experiments simulate the signal model \eqref{eq:received_matrix} and use the above metrics to evaluate the performance of the proposed set of precoders. We assume 30-antenna uniform linear arrays (ULA) \cite{tsai2004ber} with an inter-antenna spacing of half wavelength at the Tx and Rx. The Tx has a fully digital architecture and the Rx equips an array of subarrays \cite{mendez2016hybrid} with 5 RF chains. Since this work did not deal with the design of the combining matrix, we implement a simple strategy which consists in associating each RF chain to a combiner with maximum gain towards equispaced directions within the prior range of angles at the Rx. This is achieved by matching each combining vector to the array response in that direction, i.e.,
		\begin{equation} \label{eq:combiners}
		\mathbf{W} =
		\begin{pmatrix}
		\mathbf{a}_\mathrm{Rx}(\tilde{\phi}_1) & & \text{\Large 0}\\
		& \ddots \\
		\text{\Large 0} & & \mathbf{a}_\mathrm{Rx}(\tilde{\phi}_L)
		\end{pmatrix},
		\end{equation}
		where $\tilde{\phi}_1,\ldots,\tilde{\phi}_L$ are the equispaced directions within the interval $\mathcal{R}_\mathrm{Rx}$, and $\mathbf{a}_\mathrm{Rx}(\phi)\in\mathbb{C}^{\frac{N_\mathrm{Rx}}{L} \times 1}$ because each RF chain is routed to $\frac{N_\mathrm{Rx}}{L}$ antennas only.
		The SNR \eqref{eq:SNR} is fixed to \SI{-5}{\decibel}. The number of transmitted training sequences, each precoded differently, is $M=5$. The correlation parameter, defined in Section~\ref{sec:ambiguities}, which ensures identifiability of the AoD is set to $\rho=0.6$ for the AoD- and AoD-AoA-optimal precoders.
	The prior ranges on the AoD and AoA are, both, $\mathcal{R}\triangleq\mathcal{R}_\mathrm{Tx}=\mathcal{R}_\mathrm{Rx}=[\SI{90}{\degree},\SI{100}{\degree}]$.

	\subsection{Impact of Grid Size and Number of Precoders} \label{numerical_results:optimality}
	
	In Section~\ref{sec:recovery_precoders}, the solution to Problems~\ref{pro:problem:AODnAOA}--\ref{pro:problem:AOA} was obtained by approximating the prior ranges on the AoD/AoA with a grid of discrete angles. According to the analysis in Section~\ref{sec:recovery_precoders} if the grid approximation is sufficiently dense and $\rank \mathbf{X}<M$, where $\mathbf{X}$ is the solution to the convex problem \eqref{opt:conic}, then Algorithm~\ref{alg:optimal_precoders} can retrieve the optimal precoders $\mathbf{F}$ for Problems~\ref{pro:problem:AODnAOA}--\ref{pro:problem:AOA}. To verify that the grid is a good approximation, we have computed the optimal precoders for the three problems for different grid sizes. Naturally, we expect that denser grids will lead to better approximations. We have found that the worst case rEB, rDEB and rOEB for the AoD-AoA-, AoD- and AoA-optimal precoders of Problems~\ref{pro:problem:AODnAOA} to \ref{pro:problem:AOA}, respectively, quickly converges to a constant value as the grid becomes denser. In particular, the precoders computed with 5 grid points have a worst case accuracy (in terms of metrics \eqref{eq:metrics:AoD-AoA}--\eqref{eq:metrics:AoA}) that is only 1\% worse than that of the precoders computed with 100 grid points.
	
	Because problem \eqref{opt:conic} does not depend on $M$, the optimum DEB and/or OEB do not change as a function of $M$, provided that a feasible precoding matrix exists, i.e., $M\geq\rank\mathbf{X}$ as explained in Section~\ref{sec:recovery_precoders}.
	\begin{figure}
		\centering
		\begin{tikzpicture}
		\begin{semilogyaxis}[
		log ticks with fixed point,
		xlabel={Number of precoders ($M$)},
		ylabel={Worst case rDEB and rOEB [\si{\degree}]},
		xmax = 11, xmin=3,
		ymax= 10,
		cycle list name=myCycleList,
		legend pos=north west,
		legend entries={rDEB, AoD-optimal\\rOEB, AoA-optimal\\$\mathcal{R}=[\SI{80}{\degree},\SI{90}{\degree}]$\\$\mathcal{R}=[\SI{80}{\degree},\SI{100}{\degree}]$\\$\mathcal{R}=[\SI{80}{\degree},\SI{110}{\degree}]$\\},
		legend cell align=left,
		]
		\addlegendimage{no markers,thick}
		\addlegendimage{no markers,thick,dashed}
		\addlegendimage{mark=o,mark size=3,thick,color=lines-1}
		\addlegendimage{mark=star,mark size=3,thick,color=lines-2}
		\addlegendimage{mark=square,mark size=3,thick,color=lines-3}
		\addplot table[
		x=Ntraining,
		y=DEB_10,
		] {./Data/rankX.dat};
		\addplot table[
		x=Ntraining,
		y=DEB_20,
		] {./Data/rankX.dat};
		\addplot table[
		x=Ntraining,
		y=DEB_30,
		] {./Data/rankX.dat};
		\pgfplotsset{cycle list shift=-3},
		\addplot+[dashed,mark options={solid},mark=o] table[
		x=Ntraining,
		y=OEB_10,
		] {./Data/rankX.dat};
		\addplot+[dashed,mark options={solid},mark=star] table[
		x=Ntraining,
		y=OEB_20,
		] {./Data/rankX.dat};
		\addplot+[dashed,mark options={solid},mark=square] table[
		x=Ntraining,
		y=OEB_30,
		] {./Data/rankX.dat};
		\end{semilogyaxis}
		\end{tikzpicture}
		\caption{The total training time is fixed, and consequently the duration of each training sequence decreases inversely proportional to $M$. Both, the Tx and the Rx, equip a ULA. For simplicity, the prior range on the AoD and AoA are set equal, i.e., $\mathcal{R}\triangleq\mathcal{R}_\mathrm{Rx}=\mathcal{R}_\mathrm{Tx}$.}
		\label{fig:rankX}
	\end{figure}
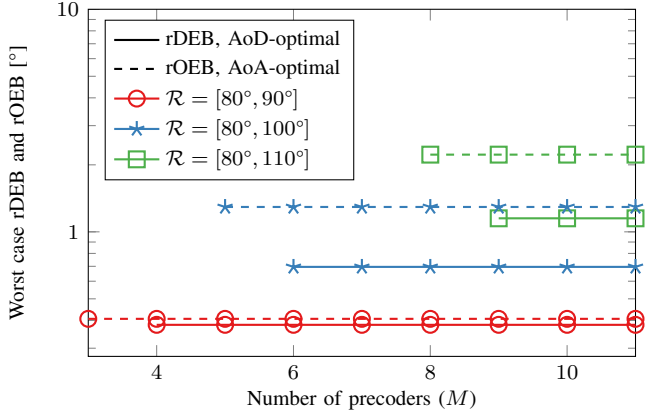
	To showcase this, Fig.~\ref{fig:rankX} plots the worst case (w.c.) rDEB and rOEB for the AoD- and AoA-optimal precoders, respectively, versus the number of precoders for different widths of the prior ranges.
	For the cases that $M<\rank\mathbf{X}$, no values of the w.c.\ rDEB and rOEB are plotted because the optimal precoders are unknown. The smallest value of $M$ in every curve is the rank of $\mathbf{X}$ (e.g., $\rank\mathbf{X}=3$ when optimizing the rOEB and $\mathcal{R}_\mathrm{Tx}=[\SI{80}{\degree},\SI{90}{\degree}]$). Therefore, the rank of $\mathbf{X}$ can be regarded as the minimum transmit diversity $M$ necessary for achieving optimal estimation accuracy. Increasing $M$ beyond this optimum value does not translate into better estimation accuracy.
	Naturally, the required Tx diversity increases with the size of the ranges of angles $\mathcal{R}$. The accuracy difference in terms of rDEB and rOEB is due to the fact that the Rx's hybrid array projects the received signal from an $N_\mathrm{Rx}$-dimensional space to an $L$-dimensional subspace through a non-optimal combining matrix, hence, lowering the quality of the observations. While not plotted here, we have observed that the rDEB and rOEB have similar accuracy when the Rx array is fully digital.

	\subsection{Estimators and Ambiguities} \label{sec:simulation_estimators}
	
	Let $\hat{\theta}$ be an estimate on the AoD. Then, the worst case root mean square error on the AoD is defined as
	\begin{equation}
	\text{Worst case rMSE} = \max_{(\theta,\phi)\in\mathcal{R}_{\mathrm{Tx}}\times\mathcal{R}_{\mathrm{Rx}}} \sqrt{\E \left(\hat{\theta}-\theta\right)^2}.
	\end{equation}
	If the MSE is tight to the DEB, such as is the case of the ML estimator, then the w.c.\ rMSE is tight to the w.c.\ rDEB. 
	To evaluate its tightness and the effect of the correlation parameter $\rho$ (introduced in Section~\ref{sec:ambiguities}), we run a Monte Carlo simulation with 100 random experiments. For each experiment we generate a random received signal \eqref{eq:received_matrix} where the channel phase is distributed randomly over the interval $[0,2\pi]$, the noise samples are drawn from a complex Gaussian random variable with variance $\sigma^2$, and the AoD/AoA are drawn from a uniform distribution over the range $\mathcal{R}=[\SI{90}{\degree},\SI{100}{\degree}]$. The channel gain and training sequence energy are fixed to $|\alpha|=\|\mathbf{s}\|_2=1$, and consequently, in order to obtain a \SI{-5}{\decibel} SNR at the Rx, the noise variance is fixed to $\sigma^2=\mathrm{SNR}^{-1}$.
	
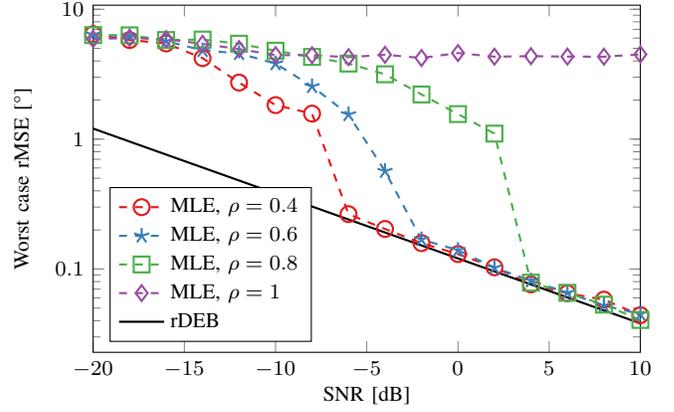
\begin{figure}
		\centering
		\begin{tikzpicture}
		\begin{semilogyaxis}[
		xlabel={SNR [\si{\decibel}]},
		ylabel={Worst case rMSE [\si{\degree}]},
		log ticks with fixed point,
		xmin=-20,xmax=10,
		cycle list name=myCycleList,
		legend pos=south west,
		legend entries = {MLE, $\rho=0.4$\\MLE, $\rho=0.6$\\MLE, $\rho=0.8$\\MLE, $\rho=1$\\rDEB\\},
		legend cell align=left,
		]
		\addplot+[dashed,mark options={solid}] table[
		x=SNR,
		y=MLE.4,
		] {./Data/variance.dat};
		\addplot+[dashed,mark options={solid}] table[
		x=SNR,
		y=MLE.6,
		] {./Data/variance.dat};
		\addplot+[dashed,mark options={solid}] table[
		x=SNR,
		y=MLE.8,
		] {./Data/variance.dat};
		\addplot+[dashed,mark options={solid}] table[
		x=SNR,
		y=MLE1,
		] {./Data/variance.dat};
		\addplot+[mark=none,black] table[
		x=SNR,
		y=DEB,
		] {./Data/variance.dat};
		\end{semilogyaxis}
		\end{tikzpicture}
		\caption{AoD estimation accuracy vs.\ SNR for different values of $\rho$ (defined in Section~\ref{sec:ambiguities}).  $\mathcal{R}_\mathrm{Rx}=\mathcal{R}_\mathrm{Tx}=[\SI{90}{\degree},\SI{100}{\degree}]$.}
		\label{fig:ambiguitiesVSsnr}
	\end{figure}

	Fig.~\ref{fig:ambiguitiesVSsnr} plots the w.c.\ rMSE for the MLE\footnote{The MLE requires maximizing the likelihood function \eqref{eq:likelihood} over the unknown parameters $(\alpha,\theta,\phi)$. The ML estimate of the channel gain requires solving a least squares whose solution can be computed analytically as a function of $\theta$ and $\phi$. Then, the ML estimate of the channel gain is plugged back to the log-likelihood function, and perform en exhaustive search over the two remaining variables $(\theta,\phi)$ using a discrete grid.} for different values of $\rho$. 
	As expected, the MLE's w.c.\ rMSE converges to the w.c.\ rDEB for a sufficient large SNR. For decreasing values of $\rho$, the SNR threshold (which is the SNR value at which the estimators' accuracy matches the lower bound) is shifted to the left. For $\rho=1$, the identifiability constraint is removed and the MLE fails to always correctly estimate the AoD for any given SNR.
	Coarsely speaking, $\rho$ controls the degree of maximum \textit{similarity} between the received signals for any two pair of AoD, where $\rho=0$ imposes signal orthogonality and $\rho=1$ removes the constraint. Thus, as $\rho$ decreases from 1 to 0, the received signals for any two AoD become less similar and more resilient to noise.

	\subsection{Comparison with Traditional Beams} \label{sec:simulation_sectored}

	Traditionally, in mmWave channel estimation, the precoders and combiners are designed such that they steer energy towards few directions.
	The channel is sounded sequentially in time with all possible pairs of precoders and combiners, and by detecting the pair leading to the largest power at the Rx, one can estimate the AoD and/or AoA with beamwidth accuracy. Its widespread use is due to the simplicity of the approach, despite not necessary being optimal in the sense of minimizing the mean error. Two of the most common approaches are sector beams \cite{kokshoorn2015fast,alkhateeb2014channel,andrews2017modeling} and maximum-gain beams \cite{han2015large,wang2009beamforming,garcia2017transmitter}, which we refer to ``sectors'' and ``beams'', respectively, for short. Since this work does not deal with the design of optimal combiners, we compare the proposed optimal precoders with sectors and beams at the Tx only.
	Sectors split the ranges $\mathcal{R}_\mathrm{Tx}$, evenly, into $M$ subregions; each sector has a large constant array gain within one subregion and low array gain outside. Beams maximize the energy towards each direction without seeking to create a flat beampattern within each subrange.
	To design the sectors we use classical tools from filter design \cite[Chapter 1.5.1]{manolakis2000statistical}, whereas the beams are designed by matching the precoder to the steering vector in that direction.
	
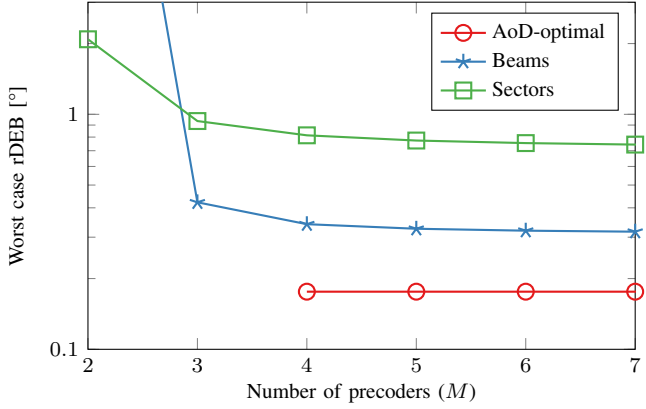
\begin{figure}
		\centering
		\begin{tikzpicture}
		\begin{semilogyaxis}[
		xlabel={Number of precoders ($M$)},
		ylabel={Worst case rDEB [\si{\degree}]},
		log ticks with fixed point,
		ymin=.1, ymax=3,
		xmin=2,xmax=7,
		cycle list name=myCycleList,
		legend pos=north east,
		legend entries = {AoD-optimal, Beams, Sectors, $N_\mathrm{Tx}=30$, $N_\mathrm{Tx}=50$},
		legend cell align=left,
		]
		\addplot table[
			x=Npilots,
			y=Opt_30,
		] {./Data/precodersComparison.dat};
		\addplot table[
			x=Npilots,
			y=Beams_30,
		] {./Data/precodersComparison.dat};
		\addplot table[
			x=Npilots,
			y=Sector_30,
		] {./Data/precodersComparison.dat};
		\end{semilogyaxis}
		\end{tikzpicture}
		\caption{Performance comparison between AoD-optimal precoders, beams and sectors. $\mathcal{R}_\mathrm{Rx}=\mathcal{R}_\mathrm{Tx}=[\SI{90}{\degree},\SI{100}{\degree}]$.}
		\label{fig:sector}
	\end{figure}

	Fig~\ref{fig:sector} plots the w.c.\ rDEB for the AoD-optimal precoders, beams and sectors.
	As outlined in Section~\ref{numerical_results:optimality}, the accuracy of the AoD-optimal precoders remains constant for any $M$ larger than $\rank\mathbf{X}$, which in this particular case is 4.
	The AoD-optimal precoders is two to four times more accurate than the beams and sectors, respectively.

	\subsection{AoA Versus AoD Estimation}
	
		A common belief is that when estimating the angle of a path at an array, operating as a receiver and estimating the AoA of the incoming path is more precise than operating as a transmitter and estimating the AoD of the outgoing path.
		For instance, in \cite{abu2017error} the authors argue that if a base station has more antennas than the user, then for estimating the direction from the base station to the user, uplink AoA estimation is more accurate than downlink AoD estimation. 
		\begin{figure}
			\centering
			\begin{tikzpicture}
			\begin{semilogyaxis}[
			xlabel={Number of antennas at the Tx ($N_\mathrm{Rx}=60-N_\mathrm{Tx}$)},
			ylabel={Worst case rEB, rDEB \& rOEB [\si{\degree}]},
			log ticks with fixed point,
			xmin=10,xmax=50,
			ymax=1.4,
			cycle list name=myCycleList,
			legend pos=north west,
			yminorticks=true,
			legend entries = {W.c.\ rDEB, AoD-optimal prec.\\W.c.\ rOEB, AoA-optimal prec.\\W.c.\ rEB, AoD-AoA-optimal\\Digital Rx\\Hybrid Rx\\},
			legend cell align=left,
			legend columns=3, 
			legend style={
				/tikz/column 2/.style={
					column sep=10pt,
				},
			},
			transpose legend
			]
			\addlegendimage{no markers,very thick,color=lines-1}
			\addlegendimage{no markers,very thick,color=lines-2}
			\addlegendimage{no markers,very thick,color=lines-3}
			\addlegendimage{no markers,very thick}
			\addlegendimage{no markers,very thick,dashed}
			\addplot+[no markers] table[
			x=Nantennas,
			y=DEB_dig,
			] {./Data/Nantennas.dat};
			\addplot+[no markers] table[
			x=Nantennas,
			y=OEB_dig,
			] {./Data/Nantennas.dat};
			\addplot+[no markers] table[
			x=Nantennas,
			y=EB_dig,
			] {./Data/Nantennas.dat};
			\pgfplotsset{cycle list shift=-3},
			\addplot+[no markers,dashed] table[
			x=Nantennas,
			y=DEB_hyb,
			] {./Data/Nantennas.dat};
			\addplot+[no markers,dashed] table[
			x=Nantennas,
			y=OEB_hyb,
			] {./Data/Nantennas.dat};
			\addplot+[no markers,dashed] table[
			x=Nantennas,
			y=EB_hyb,
			] {./Data/Nantennas.dat};
			\end{semilogyaxis}
			\end{tikzpicture}
			\caption{Comparison between AoD and AoA estimation. The sum of the number of antennas at the Tx and Rx are kept constant to 60 antennas; so as the number of Tx antennas increases, the number of Rx antennas decreases, and vice-versa.  $\mathcal{R}_\mathrm{Rx}=\mathcal{R}_\mathrm{Tx}=[\SI{90}{\degree},\SI{100}{\degree}]$.}
			\label{fig:Nanntenas}
		\end{figure}
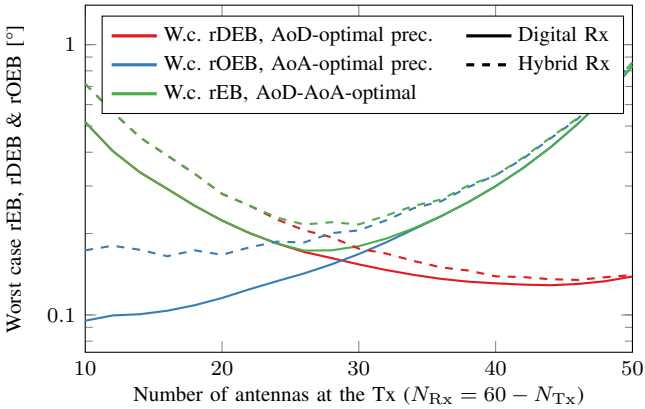
	
		Fig.~\ref{fig:Nanntenas} plots the bounds on the AoD/AoA estimation accuracy \eqref{eq:metrics:AoD-AoA}--\eqref{eq:metrics:AoA} versus the number of antennas at the Tx and Rx, assuming the total number of antennas is 60. The goal is to observe the effect of accumulating more antennas at the Tx or Rx on the estimation accuracy of the AoD and AoA. For a fair comparison, we begin by assuming that both arrays have fully digital architectures (represented as solid lines in the figure). For the case where the Tx and Rx have the same number of antennas (i.e., $N_\mathrm{Tx}=N_\mathrm{Rx}=30$), the accuracy is very similar whether estimating the AoD or AoA ($\sim\SI{0.2}{\degree}$). However, for the two extreme cases where the Rx has 50 antennas and the Tx only 10 ($\sim\SI{0.1}{\degree}$ at the left side of the figure), the AoA estimation accuracy slightly outperforms the AoD accuracy for the case the Tx has 50 antennas and the Rx only 10 ($\sim\SI{0.2}{\degree}$ at the right side of the figure).
		Jumping to the case where the Rx equips a hybrid array with 5 RF chains, we notice that the AoD and AoA estimation accuracy worsens, in particular for the AoA. This loss of accuracy caused by the hybrid Rx is expected because, mathematically, the Rx is compressing the received signal onto a $L$-dimensional subspace.
		Thus, in general, the AoD and AoA estimation accuracy will differ. However, in the particular case where both arrays have the same number of antennas, are fully digital, and the Tx performs optimal precoding, operating as a Tx and estimating the AoD, or operating as a Rx and estimating the AoA, result in similar accuracy.

	\subsection{Complexity Analysis}
	
	Obtaining the optimal precoders requires solving problem~\eqref{opt:conic}. This is a conic (hence, convex) optimization problem \cite{ben2001lectures} because the constraints are composed of second-order cones \eqref{opt:conic:DEB} and a positive semidefinte cone \eqref{opt:conic:SD}. Conic problems can be efficiently solved by the interior-points method \cite{Mosek2016}. Second-order cone programs (SOCPs) and semidefinite programs (SDPs) are particular instances of conic problems where all constraints are second-order cones or positive semidefinite cones, respectively. These categories are important because, while both types of problems have polynomial complexity, for equal number of variables and constraints, SDPs are generally substantially more computationally complex than SOCPs.
	By leveraging the Schur complement, the proposed conic problem \eqref{opt:conic} could be reformulated as an SDP by transforming all second-order cone into positive semidefinite cones. However, it would not be wise in terms of computational complexity. Unfortunately, to the best of the author's knowledge, the computational complexity of conic problems with mixed cones is not well understood in the literature, and an analysis on its SDP form would lead to too pessimistic bounds. Thus, we resolve to a numerical analysis on the computational complexity.
	
\begin{figure}
		\centering
		\begin{tikzpicture}
		\begin{axis}[
		xlabel={Number of grid points ($S_\mathrm{Tx}$)},
		ylabel={Average duration [\si{\second}]},
		xmin=2,xmax=20,
		cycle list name=myCycleList,
		legend pos=north west,
		legend entries = {$N_\mathrm{Tx}=10$,$N_\mathrm{Tx}=20$,$N_\mathrm{Tx}=30$,$N_\mathrm{Tx}=40$},
		legend cell align=left,
		]
		\addplot table[
			x=Ngrid,
			y=N_10,
		] {./Data/complexity.dat};
		\addplot table[
			x=Ngrid,
			y=N_20,
		] {./Data/complexity.dat};
		\addplot table[
			x=Ngrid,
			y=N_30,
		] {./Data/complexity.dat};
		\addplot table[
			x=Ngrid,
			y=N_40,
		] {./Data/complexity.dat};
		\end{axis}
		\end{tikzpicture}
		\caption{Average duration for obtaining the AoD-AoA-optimal precoders which involves solving problem~\eqref{opt:conic} and running Algorithm~\ref{alg:optimal_precoders}. The the two main parameters affecting the number of variables and constraints are the number of antennas at the Tx, $N_\mathrm{Tx}$, and the number of grid points $S_\mathrm{Tx}$.}
		\label{fig:complexity}
	\end{figure}
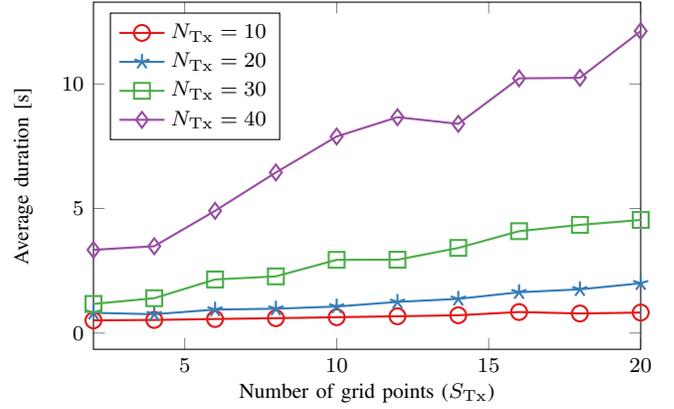

	Fig.~\ref{fig:complexity}  plots the average duration of solving problem~\eqref{opt:conic} and executing Algorithm~\ref{alg:optimal_precoders} with MATLAB in a laptop computer with \SI{2}{\giga\hertz} clock speed. 
	From the figure we observe that the duration appears to be linear with the number of grid points and superlinear with the number of antennas.

	\section{Conclusions} \label{sec:conclusions}
	
	This paper dealt with the design of precoders for mmWave communication, optimized for estimating the AoD and AoA under a given uncertainty range. Our focus was on a single propagation path, assuming that such a path has been identified for precoding and combining so that weaker paths can be ignored. The design is based on minimizing the worst-case CRB of the AoD and AoA, which is a tight lower bound to the variance of MLE at medium-to-high SNRs. Through a convex reformulation, optimal precoders are recovered, leading to a two-fold or larger improvement on the estimation accuracy with respect to traditional schemes. We found that beyond a certain number of precoders at the Tx, the CRB on the AoD/AoA does not improve and that the minimum number of precoders for optimal performance is a side-product of the optimization procedure. Finally, numerical evidence shows an array can estimate the direction of a path in transmit (AoD) or receive (AoA) mode with similar accuracy when performing optimal precoding.

	\appendices
	
	\section{Fisher Information Matrix and Cram\'{e}r-Rao Bound} \label{sec:CRB}
	
	Assume perfect timing recovery, then the log-likelihood function of the observations \eqref{eq:received_matrix} after neglecting constant terms is proportional to
	\begin{equation} \label{eq:likelihood}
	\log \rho\left(\mathbf{Y} | \theta, \phi, \alpha  \right) \propto
	-\frac{1}{\sigma^2}\left\|\mathbf{Y} - \alpha
	\|\mathbf{s}\|_2
	\mathbf{b}_\mathrm{Rx}\left(\phi\right) \mathbf{a}_\mathrm{Tx}^\mathrm{H}(\theta) \mathbf{F} \right\|_\mathrm{F}^2.
	\end{equation}
	According to the analysis of \cite{shahmansoori5g}, the Fisher information matrix of the unknown parameters turns out to be
	\begin{equation}
	\mathbf{J} =
	\begin{bmatrix}
	\Phi_{\theta,\theta}	& \Phi_{\theta,\phi} & \Phi_{\theta,\Re\alpha} & \Phi_{\theta,\Im\alpha} \\
	\Phi_{\theta,\phi}	& \Phi_{\phi,\phi} & \Phi_{\phi,\Re\alpha} & \Phi_{\phi,\Im\alpha} \\
	\Phi_{\theta,\Re\alpha}	& \Phi_{\phi,\Re\alpha} & \Phi_{\Re\alpha,\Re\alpha} & 0												   \\
	\Phi_{\theta,\Im\alpha}	& \Phi_{\phi,\Im\alpha} & 0												  & \Phi_{\Im\alpha,\Im\alpha}
	\end{bmatrix},
	\end{equation}
	where\footnote{Contrary to \cite{shahmansoori5g}, we have not assumed that $\|\mathbf{a}_\mathrm{Tx}(\theta)\|={\|\mathbf{a}_\mathrm{Rx}(\phi)\|=1}$.}
	\begingroup
	\allowdisplaybreaks
	\begin{subequations}
		\begin{align}
		\Phi_{\theta,\theta} &=
		2\, \mathrm{SNR} \left\|\mathbf{b}_\mathrm{Rx}(\phi)\right\|_2^2 \left\| \dot{\mathbf{b}}_\mathrm{Tx}(\theta) \right\|_2^2 \\
		\Phi_{\phi,\phi} &=
		2\, \mathrm{SNR} \left\|\dot{\mathbf{b}}_\mathrm{Rx}(\phi)\right\|_2^2 \left\|  \mathbf{b}_\mathrm{Tx}(\theta) \right\|_2^2 \\
		\Phi_{\Re\alpha,\Re\alpha} &= \Phi_{\Im\alpha,\Im\alpha} = \frac{2\|s\|_2^2}{\sigma^2} \left\|\mathbf{b}_\mathrm{Rx}(\phi)\right\|_2^2 \left\| \mathbf{b}_\mathrm{Tx}(\theta) \right\|_2^2 \\
		\Phi_{\theta,\phi} &= 
		2\, \mathrm{SNR} \,\Re \left[\mathbf{b}_\mathrm{Rx}^\mathrm{H}(\phi)\dot{\mathbf{b}}_\mathrm{Rx}(\phi) \mathbf{b}_\mathrm{Tx}^\mathrm{H} (\theta)  \dot{\mathbf{b}}_\mathrm{Tx}(\theta)\right] \\
		\Phi_{\theta,\Re\alpha} &=
		\frac{2\|s\|_2^2}{\sigma^2}\left\|\mathbf{b}_\mathrm{Rx}(\phi)\right\|_2^2 \Re \left[ \alpha \, \dot{\mathbf{b}}_\mathrm{Tx}^\mathrm{H} (\theta)  \mathbf{b}_\mathrm{Tx}(\theta)\right] \\
		\Phi_{\theta,\Im\alpha} &=
		\frac{2\|s\|_2^2}{\sigma^2} \left\|\mathbf{b}_\mathrm{Rx}(\phi)\right\|_2^2 \Im \left[ \alpha \, \dot{\mathbf{b}}_\mathrm{Tx}^\mathrm{H} (\theta) \mathbf{b}_\mathrm{Tx}(\theta)\right] \\
		\Phi_{\phi,\Re\alpha} &=
		\frac{2\|s\|_2^2}{\sigma^2} \Re \left[\alpha \, \mathbf{b}_\mathrm{Rx}^\mathrm{H}(\phi) \dot{\mathbf{b}}_\mathrm{Rx}(\phi) \right] \left\| \mathbf{b}_\mathrm{Tx}(\theta) \right\|_2^2 \\
		\Phi_{\phi,\Im\alpha} &=
		\frac{2\|s\|_2^2}{\sigma^2} \Im \left[ \alpha \, \mathbf{b}_\mathrm{Rx}^\mathrm{H}(\phi) \dot{\mathbf{b}}_\mathrm{Rx}(\phi) \right] \left\| \mathbf{b}_\mathrm{Tx}(\theta) \right\|_2^2,
		\end{align}
	\end{subequations}
	$\mathbf{b}_\mathrm{Tx}(\theta) \triangleq \mathbf{F}^\mathrm{H} \mathbf{a}_\mathrm{Tx}(\theta)$, $\dot{\mathbf{b}}_\mathrm{Tx}(\theta) \triangleq \frac{\text{d}}{\text{d}\theta}\mathbf{F}^\mathrm{H}\mathbf{a}_\mathrm{Tx}(\theta)$ and $\mathrm{SNR}=|\alpha|^2 \|\mathbf{s}\|_2^2 \sigma^{-2}$.
	The CRB requires inverting the FIM. Define
	\begin{align}
	\mathbf{J}_{11} &=
	\begin{bmatrix}
	\Phi_{\theta,\theta}	& \Phi_{\theta,\phi} \\
	\Phi_{\theta,\phi}	& \Phi_{\phi,\phi}
	\end{bmatrix} \\
	\mathbf{J}_{12} = \mathbf{J}_{21}^T &=
	\begin{bmatrix}
	\Phi_{\theta,\Re\alpha} & \Phi_{\theta,\Im\alpha} \\
	\Phi_{\phi,\Re\alpha} & \Phi_{\phi,\Im\alpha}
	\end{bmatrix} \\
	\mathbf{J}_{22} &=
	\begin{bmatrix}
	\Phi_{\Re\alpha,\Re\alpha} & 0												   \\
	0												  & \Phi_{\Im\alpha,\Im\alpha}
	\end{bmatrix}.
	\end{align}
	Then, by the block matrix inversion formula, the CRB on the AoD and AoA is
	\begin{equation}
	\left(\mathbf{J}_{11}-\mathbf{J}_{12}\mathbf{J}_{22}^{-1}\mathbf{J}_{21}\right)^{-1} =
	\begin{bmatrix}
	\mathrm{DEB} & 0 \\
	0 & \mathrm{OEB}
	\end{bmatrix}
	\end{equation}
	where
	\begin{align}
	\mathrm{DEB}^{-1} &= {\textstyle 2\, \mathrm{SNR} \left\|\mathbf{b}_\mathrm{Rx}(\phi)\right\|_2^2
		\left(
		\left\| \dot{\mathbf{b}}_\mathrm{Tx}(\theta) \right\|_2^2
		-\frac{\left|\tilde{\mathbf{a}}_\mathrm{Tx}^\mathrm{H} (\theta) \dot{\mathbf{b}}_\mathrm{Tx}(\theta)\right|^2}{  \left\| \mathbf{b}_\mathrm{Tx}(\theta) \right\|_2^2}
		\right)} \label{eq:DEB} \\
	\mathrm{OEB}^{-1} &= {\textstyle 2\, \mathrm{SNR} \left\| \mathbf{b}_\mathrm{Tx}(\theta) \right\|_2^2
		\left(
		\left\|\dot{\mathbf{b}}_\mathrm{Rx}(\phi)\right\|_2^2
		-\frac{\left|\mathbf{b}_\mathrm{Rx}^\mathrm{H}(\phi)\dot{\mathbf{b}}_\mathrm{Rx}(\phi)\right|^2 }{\left\|\mathbf{b}_\mathrm{Rx}(\phi)\right\|_2^2 }
		\right)}. \label{eq:OEB}
	\end{align}
	When computing the DEB (or OEB) for uniform linear arrays (ULA), to avoid dealing with zero-information points \cite{bashan2007estimation}, we recommend the use of the `spatial frequency' $\omega\triangleq \cos(\theta)/2$ (or $\omega\triangleq \cos(\phi)/2$) as unknown parameter instead of $\theta$ (or $\phi$) as done in \cite{stoica1989music}.

	\section{Interpretation of the DEB and OEB} \label{sec:interpretation}
	
	The intuition behind DEB's expression \eqref{eq:bound:AoD}, \eqref{eq:DEB} is a bit cumbersome since it includes multiple terms that depend on $\mathbf{F}$.
	For notation convenience, define $\mathbf{b}_\mathrm{Tx}(\theta,\mathbf{F}) \triangleq \mathbf{F}^\mathrm{H} \mathbf{a}_\mathrm{Tx}(\theta)$ and its derivative $\dot{\mathbf{b}}_\mathrm{Tx}(\theta,\mathbf{F}) \triangleq \frac{\mathrm{d}}{\mathrm{d}\theta}\mathbf{b}_\mathrm{Tx}(\theta,\mathbf{F})$.
	By omitting all factors that do not depend on $\mathbf{F}$, the DEB may be expressed as
	\begin{align}
	\mathrm{DEB} &\propto
	\left(
	\left\|\dot{\mathbf{b}}_\mathrm{Tx}(\theta,\mathbf{F})\right\|_2^2
	-\left|
	\frac{\mathbf{b}_\mathrm{Tx}^\mathrm{H}(\theta,\mathbf{F})}{\left\|\mathbf{b}_\mathrm{Tx}(\theta,\mathbf{F})\right\|_2}
	\dot{\mathbf{b}}_\mathrm{Tx}(\theta,\mathbf{F})
	\right|^2
	\right)^{-1} \\
	&= \left\|\dot{\mathbf{b}}_\mathrm{Tx}^\perp(\theta,\mathbf{F})\right\|_2^{-2},
	\end{align}
	where $\dot{\mathbf{b}}_\mathrm{Tx}^\perp(\theta,\mathbf{F})$ is the projection of $\dot{\mathbf{b}}_\mathrm{Tx}(\theta,\mathbf{F})$ onto the orthogonal complement of $\mathbf{b}_\mathrm{Tx}(\theta,\mathbf{F})$.
	The function $\mathbf{b}_\mathrm{Tx}(\theta,\mathbf{F})$ not only appears in the CRB, but careful attention reveals that it also appears in the signal model \eqref{eq:received_matrix},
	\begin{equation}
	\mathbf{Y} 
	= \alpha \|\mathbf{s}\|_2 \mathbf{b}_\mathrm{Rx}(\phi) \mathbf{b}_\mathrm{Tx}^\mathrm{H}(\theta,\mathbf{F}) + \mathbf{N},
	\end{equation}
	and it carries all information about the AoD.
	Since the CRB is a tight lower bound on the variance when the noise tends to zero, the signal $\mathbf{b}_\mathrm{Tx}(\theta,\mathbf{F})$ should be analyzed for the case where a small perturbation is applied to the AoD, in which case a good approximation is its first order Taylor polynomial
	\begin{align} \label{eq:Taylor_signature}
	\mathbf{b}_\mathrm{Tx}(\theta',\mathbf{F}) \approx \mathbf{b}_\mathrm{Tx}(\theta,\mathbf{F}) + (\theta'-\theta) \dot{\mathbf{b}}_\mathrm{Tx}(\theta,\mathbf{F}),
	\end{align}
	where $\theta'$ and $\theta$ represent two closely spaced AoD. 
	Let $\beta \mathbf{b}_\mathrm{Tx}(\theta,\mathbf{F})$ be the projection of $\dot{\mathbf{b}}_\mathrm{Tx}(\theta,\mathbf{F})$ onto $\mathbf{b}_\mathrm{Tx}(\theta,\mathbf{F})$, then,
	\begin{equation} \label{eq:Taylor_signature_decomp}
	\mathbf{b}_\mathrm{Tx}(\theta',\mathbf{F}) \approx \left[1+(\theta'-\theta)\beta\right]\mathbf{b}_\mathrm{Tx}(\theta,\mathbf{F}) + (\theta'-\theta) \dot{\mathbf{b}}_\mathrm{Tx}^\perp(\theta,\mathbf{F}).
	\end{equation}
	From \eqref{eq:Taylor_signature} we can infer that if $\dot{\mathbf{b}}_\mathrm{Tx}(\theta,\mathbf{F})$ is large, $\mathbf{b}_\mathrm{Tx}(\theta',\mathbf{F})$ and $\mathbf{b}_\mathrm{Tx}(\theta,\mathbf{F})$ will differ more. However, if $\dot{\mathbf{b}}_\mathrm{Tx}(\theta,\mathbf{F})$ and $\mathbf{b}_\mathrm{Tx}(\theta,\mathbf{F})$ are collinear (only differ by a scaling factor), then $\dot{\mathbf{b}}_\mathrm{Tx}^\perp(\theta,\mathbf{F})=0$, and by \eqref{eq:Taylor_signature_decomp} the two signals will also be collinear. Since the channel gains are unknown, two collinear signals will be indistinguishable at the Rx, making the AoD unidentifiable. Thus, as asserted by the DEB, the accuracy in estimating the AoD depends on the magnitude of the part of $\dot{\mathbf{b}}_\mathrm{Tx}(\theta,\mathbf{F})$ orthogonal to $\mathbf{b}_\mathrm{Tx}(\theta,\mathbf{F})$.

	Regarding the OEB \eqref{eq:bound:AoA}, \eqref{eq:OEB}, from the Tx's perspective, its minimization boils down to maximizing $\|\mathbf{F}^\mathrm{H} \mathbf{a}_\mathrm{Tx}(\theta)\|_2^2$, i.e., maximizing the aggregated array gain \eqref{eq:total_gain} towards direction $\theta$, which makes sense because maximizing the received energy helps combat the noise.

	\bibliographystyle{IEEEtran}
	\bibliography{IEEEabrv,./references}
	
\end{document}